\documentclass[structabstract]{aa}  

\usepackage{amstext,amsmath}
\usepackage{graphicx}
\usepackage{txfonts}	
\usepackage{natbib}

\begin{document}

     \title{Stable Umbral Chromospheric Structures} 

   \author{V. M. J.  Henriques%
          \inst{1}
	E.  Scullion%
          \inst{2}
          	M.  Mathioudakis%
          \inst{1}
          D. Kiselman 
\inst{3}         
 P. T. Gallagher
\inst{2}
          	F. P. Keenan%
          \inst{1}
          }

   \institute{Astrophysics Research Centre, School of Mathematics and Physics, Queen's University, Belfast BT7 1NN, Northern Ireland, UK: \email{v.henriques@qub.ac.uk}
          \and 
		School of Physics, Trinity College Dublin, Dublin 2, Ireland:\email{scullie@tcd.ie}
          \and
             Institute for Solar Physics, Dept. of Astronomy, Stockholm University, Albanova University Center, 106 91 Stockholm, Sweden\\
                         }

\offprints{V. M. J. Henriques, \email{v.henriques@qub.ac.uk}}
                                                   
      \date{Submitted July 23, 2014; Accepted December 16, 2014}
  
  \abstract{}{To understand the morphology of the chromosphere in sunspot umbra. We investigate if the horizontal structures observed in the spectral core of the \ion{Ca}{ii}~H line are ephemeral visuals caused by the shock dynamics of more stable structures, and examine their relationship with observables in the H-alpha line.}{Filtergrams in the core of the \ion{Ca}{ii}~H  and H-alpha lines as observed with the Swedish 1-m Solar Telescope are employed. We utilise a technique that creates composite images and tracks the flash propagation horizontally.}{We find 0\farcs 15 wide horizontal structures, in all of the three target sunspots, for every flash where the seeing was moderate to good. Discrete dark structures are identified that are stable for at least two umbral flashes, as well as systems of structures that live for up to 24~minutes. We find cases of extremely extended structures with similar stability, with one such structure showing an extent of 5\arcsec. Some of these structures have a correspondence in H-alpha but we were unable to find a one to one correspondence for every occurrence. If the dark streaks are formed at the same heights as umbral flashes then there are systems of structures with strong departures from the vertical for all three analysed sunspots.}{Long-lived \ion{Ca}{ii}~H filamentary horizontal structures are a common and likely ever-present feature in the umbra of sunspots. If the magnetic field in the chromosphere of the umbra is indeed aligned with the structures, then the present theoretical understanding of the typical umbra needs to be revisited.} 


\authorrunning{V. M. J. Henriques}
\titlerunning{}

   \maketitle

\section{Introduction}

The magnetic field topology in the chromospheric umbra of sunspots is generally thought to be uniformly vertical. This assumption stems from the polarimetric observations of umbrae invariably finding the field inclination to be very close to vertical, at least in its central parts  \citep{2003A&ARv..11..153S,2011LRSP....8....4B}. Radiative transfer calculations of vertical simulated model umbra atmospheres, with pre-set fields and a piston \citep{2010ApJ...722..888B}, accurately reproduce the saw toothed-like pattern of alternating blue and red-shifted emitting features, which are observed in the spectra of \ion{Ca}{ii}~H umbral flashes \citep{2003A&A...403..277R,2014ApJ...786..137T}.   

However, there is evidence that significantly complicates this view of a purely vertical magnetic field. At the photospheric level we now know that opposite polarity patches exist even in the innermost penumbra \citep{2013A&A...553A..63S} and there is strong evidence for fine horizontal structure in the sunspot chromosphere. Polarimetry in Ca II 8542\AA\  \citep{2000Sci...288.1398S,2000ApJ...544.1141S} showed abnormal profiles that could be explained by a two-component atmosphere shifted in both velocity and polarisation.  Further indirect evidence of a multi-component umbral atmosphere was obtained by \cite{2005ApJ...635..670C} and \cite{2008ApJ...686L..45T}. 

Filamentary structures with surprisingly large apparent horizontal extents, of up to 2000~km, were directly observed by \cite{2009ApJ...696.1683S} in the  \ion{Ca}{ii}~H line core using Hinode, and again at higher resolution (0\farcs 10) by \cite{2013A&A...557A...5H}. These become visible over an area when the bright component of an umbral flash is active in that area. \cite{2009ApJ...696.1683S} states that it is almost as if the flash ``illuminates" the filaments. For simplicity we will refer to these structures as UCFs (Umbral \ion{Ca}{ii}~H/K Fibrils) throughout this paper. 

Since then, in H-alpha observations of a sunspot, \cite{2013ApJ...776...56R} observed filamentary structures that have nearly identical properties to Dynamic Fibrils \cite{2006ApJ...647L..73H}, only shorter and with lower velocities.These are likely due to the lower wave power availability and the higher inclination of the wave guides. They also found similar dynamic behaviour in \ion{Ca}{ii}~8542, where the blue wings revealed up-shooting material and the red wings down-flowing material, both with parabolic trajectories, similar to those of the H-alpha fibrils. They concluded from these observations that the structures observed in \ion{Ca}{ii}~H by \cite{2009ApJ...696.1683S} and \cite{2013A&A...557A...5H}, are likely to be counterparts. However, no 8542 fibrils were discernible over the umbra. In H-alpha, with a resolution of up to 0\farcs 10 and clearly visible over a large patch of umbra,  \cite{2014ApJ...787...58Y} observed jet-like features that they call spikes, with properties that are very similar to those of  \cite{2013ApJ...776...56R}. The authors conclude that these are likely to be the same. Both \cite{2013ApJ...776...56R} and \cite{2014ApJ...787...58Y} argue that asymmetric propagation of shocks, as simulated by \cite{2011ApJ...743..142H} in a generic magnetic atmosphere, are the most likely explanation for their observations.

Also, directly over the umbra and in \ion{Ca}{ii}~H, \cite{2013A&A...552L...1B} have found bright, short lived (50s) jets which the authors call Umbral Micro-jets. These are likely different from both UCFs and short dynamic fibrils (2-3~min) due to their shorter lifetimes and lower frequency.

Other observations hinting at fine structure in the umbra include \cite{2013A&A...556A.115D}, who use non-LTE inversions in the Ca II 8542\AA\ line as well as the weak-field approximation to find spatial variations of 200-300~G in the magnetic field strength of a sunspot umbra. Even though they do not directly see fibril-like structures, the variation in magnetic field they detect suggests fine structure above what may be expected from noise. 

In a broader context, the simulations of  \cite{2010ApJ...722..888B} show that slow mode waves from the deeper layers of the umbra steepen into shocks at chromospheric heights, causing emission in the \ion{Ca}{ii}~H core, very similar to the grain formation mechanism put forward for the general atmosphere \citep{1997ApJ...481..500C}. This strengthened earlier evidence for non-linear waves and shocks such as  \cite{1997ESASP.404..189B}, \cite{2003A&A...403..277R}  and later \cite{2010ApJ...722..131F}. These shocks steepen from the 3-minute oscillations, observed at the discovery of umbral flashes themselves \citep{1969SoPh....7..351B} as well as in early follow up papers \citep{1969SoPh....7..366W}. In turn, the 3-minute oscillations are likely caused by a resonance in the umbral atmosphere itself (see early review by \cite{1985AuJPh..38..811T}) and are, at least partially, powered by the photospheric 5-min oscillations \citep{2006ApJ...640.1153C}. 

In the simulations of \cite{2010ApJ...722..888B}, the atmosphere is pre-set to have a vertical magnetic field. They find the emission peaking at heights of 1.1 to 1.3 Mm above optical depth unity at 500~nm (with an alternate simulation showing 0.9 Mm to 1.1 Mm). Most importantly, the full range of the contribution function from different heights is generally strongly peaked within a span of up to 300~km, with tails that can range by up to 1 Mm at the final stages of the simulated flashes.    

Here we investigate the topic of UCFs in three different sunspots. We focus on their stability and attempt to estimate their inclination from the vertical.  The formation ranges computed in \cite{2010ApJ...722..888B}, together with the measured projected-horizontal extend of the structures, are used to compute an interval of possible inclination angles. Furthermore, we search for evidence of both elongated and short structures corresponding to very large and very small inclinations from the vertical, respectively.

\section{Observations and data reduction}
\label{sect:setup}

\begin{figure*}[!ht]
\centering\resizebox{6.195cm}{!}{\includegraphics[clip=true]{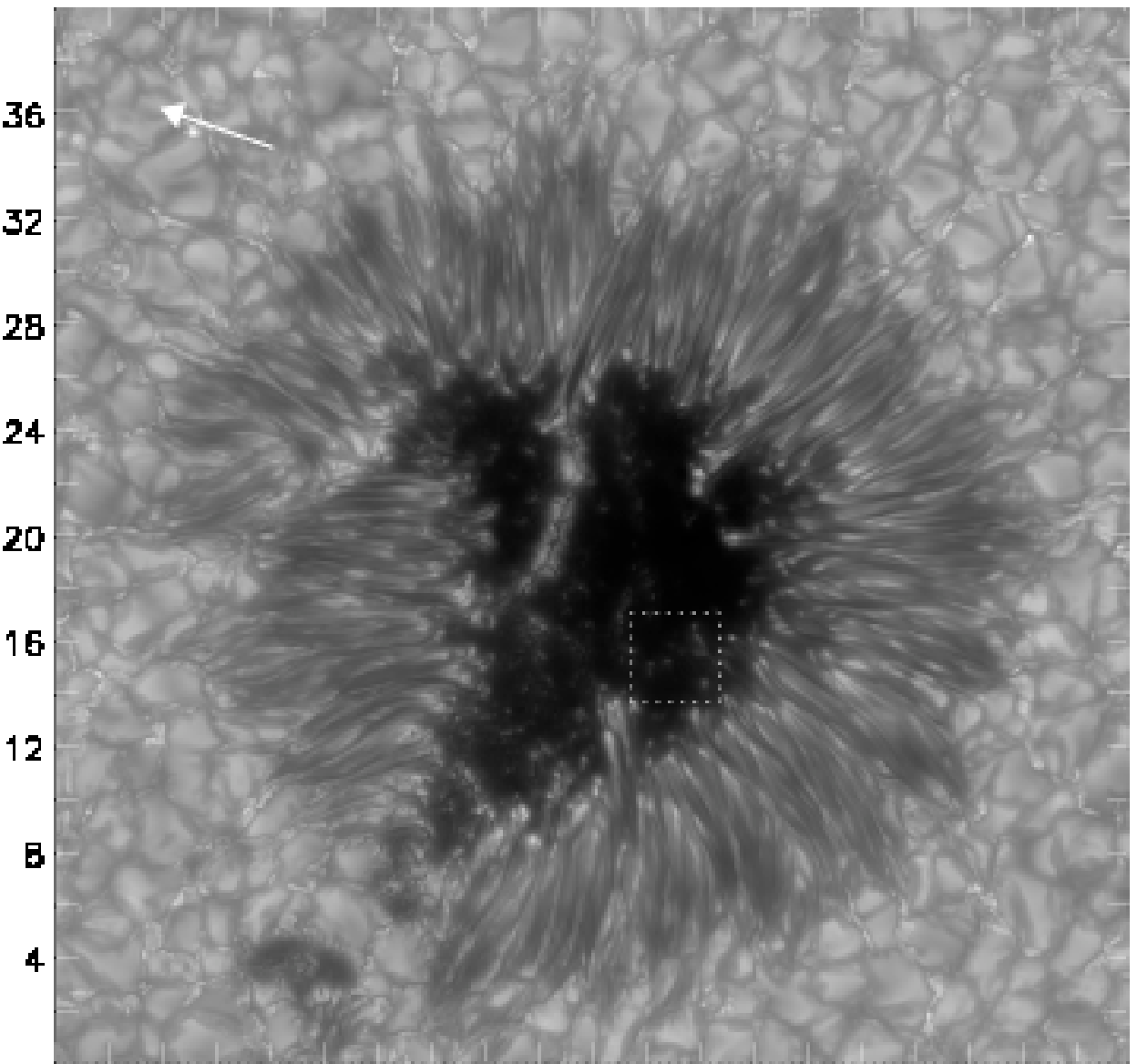}}
\centering\resizebox{5.9cm}{!}{\includegraphics[clip=true]{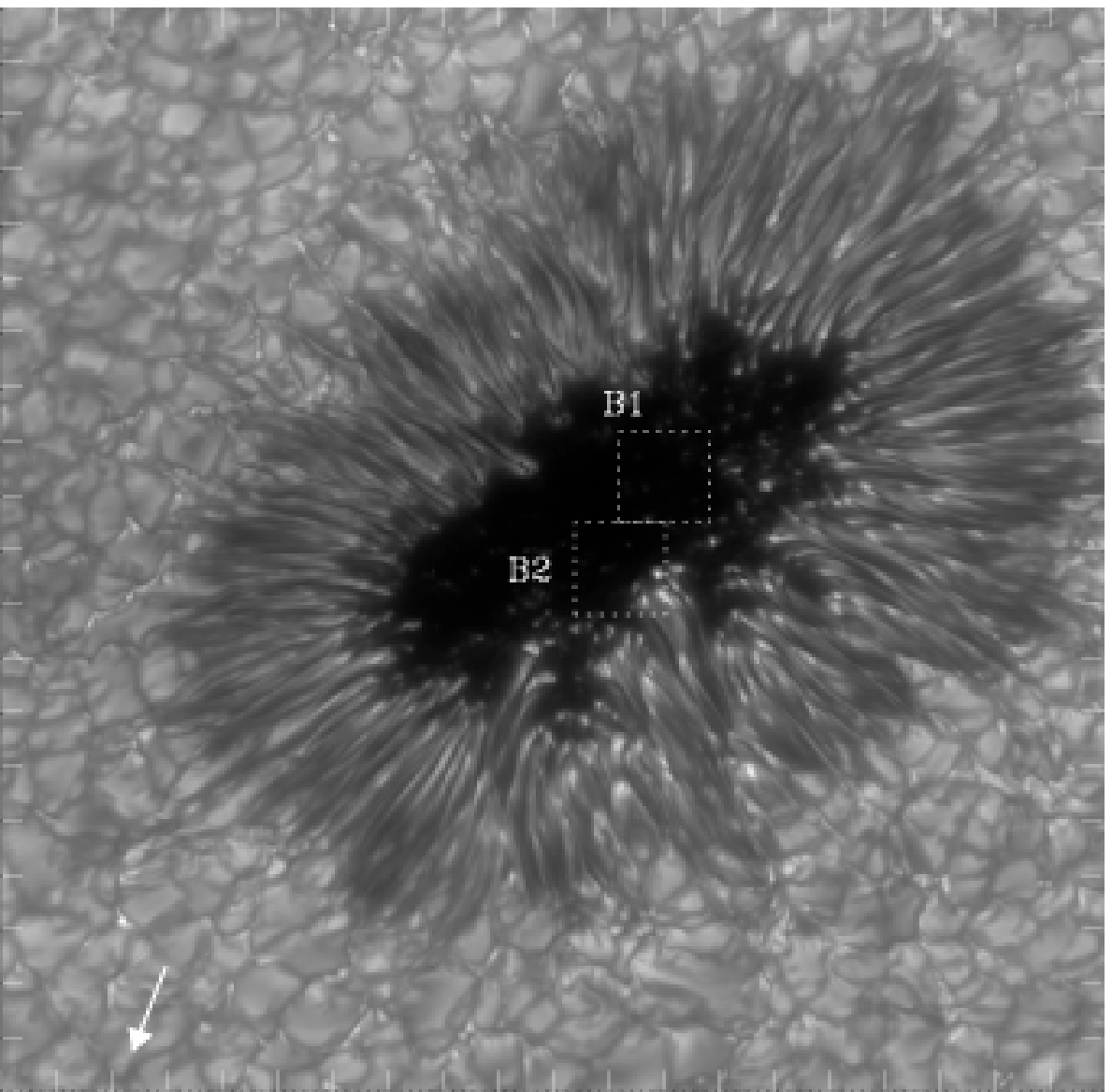}}
\centering\resizebox{5.9cm}{!}{\includegraphics[clip=true]{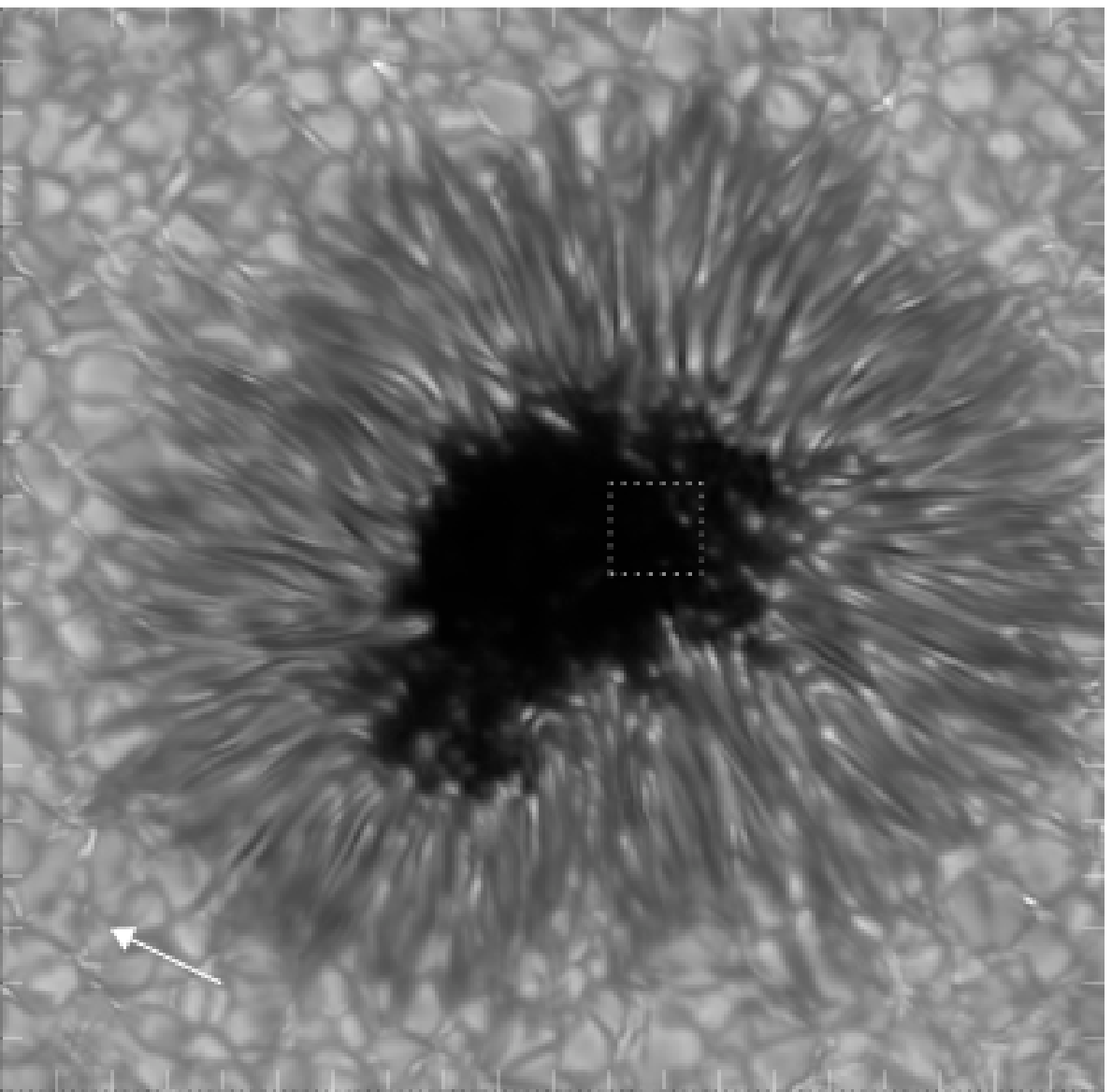}}

\resizebox{6.195cm}{!}{\includegraphics[clip=true]{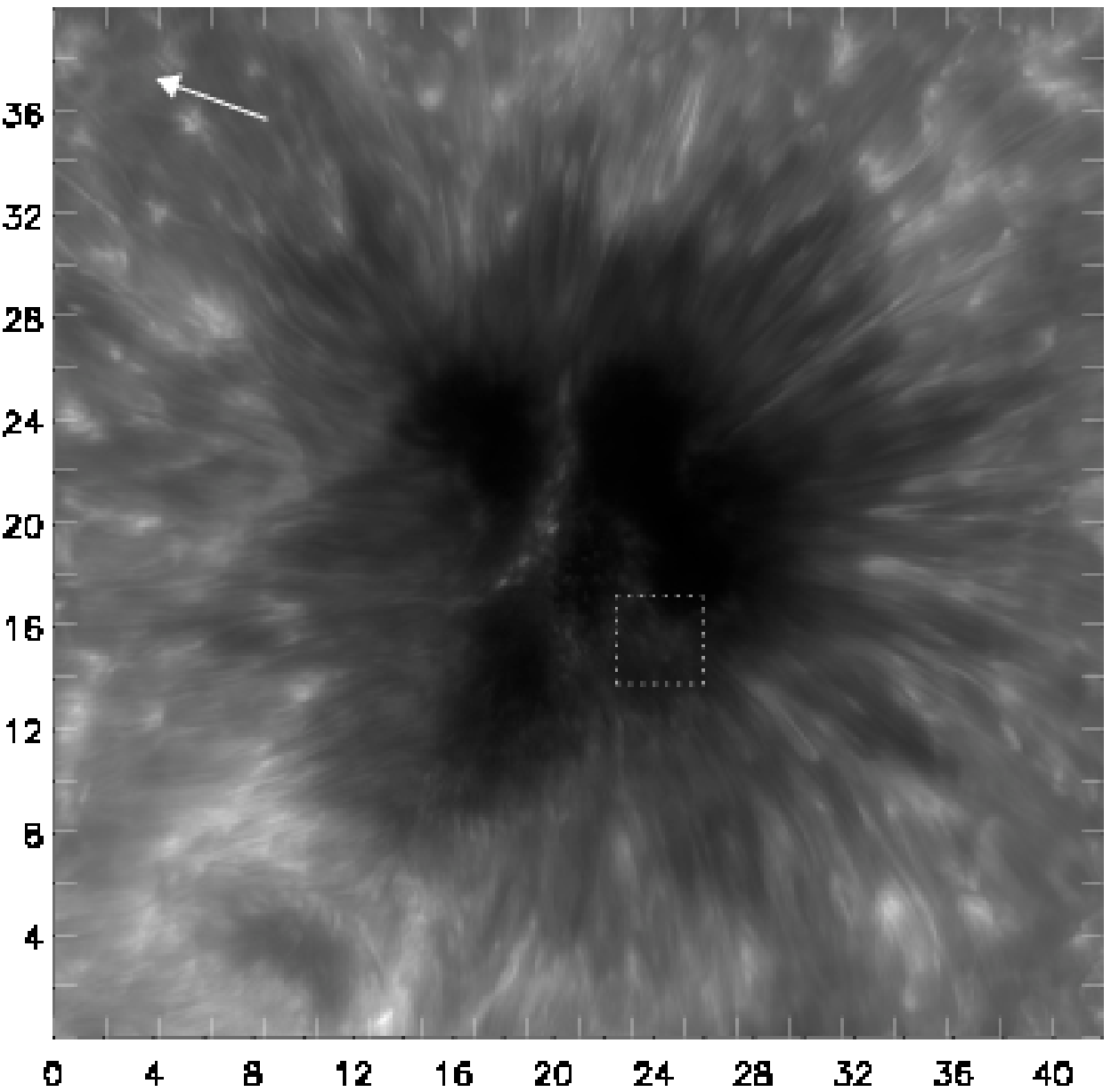}}
\resizebox{5.9cm}{!}{\includegraphics[clip=true]{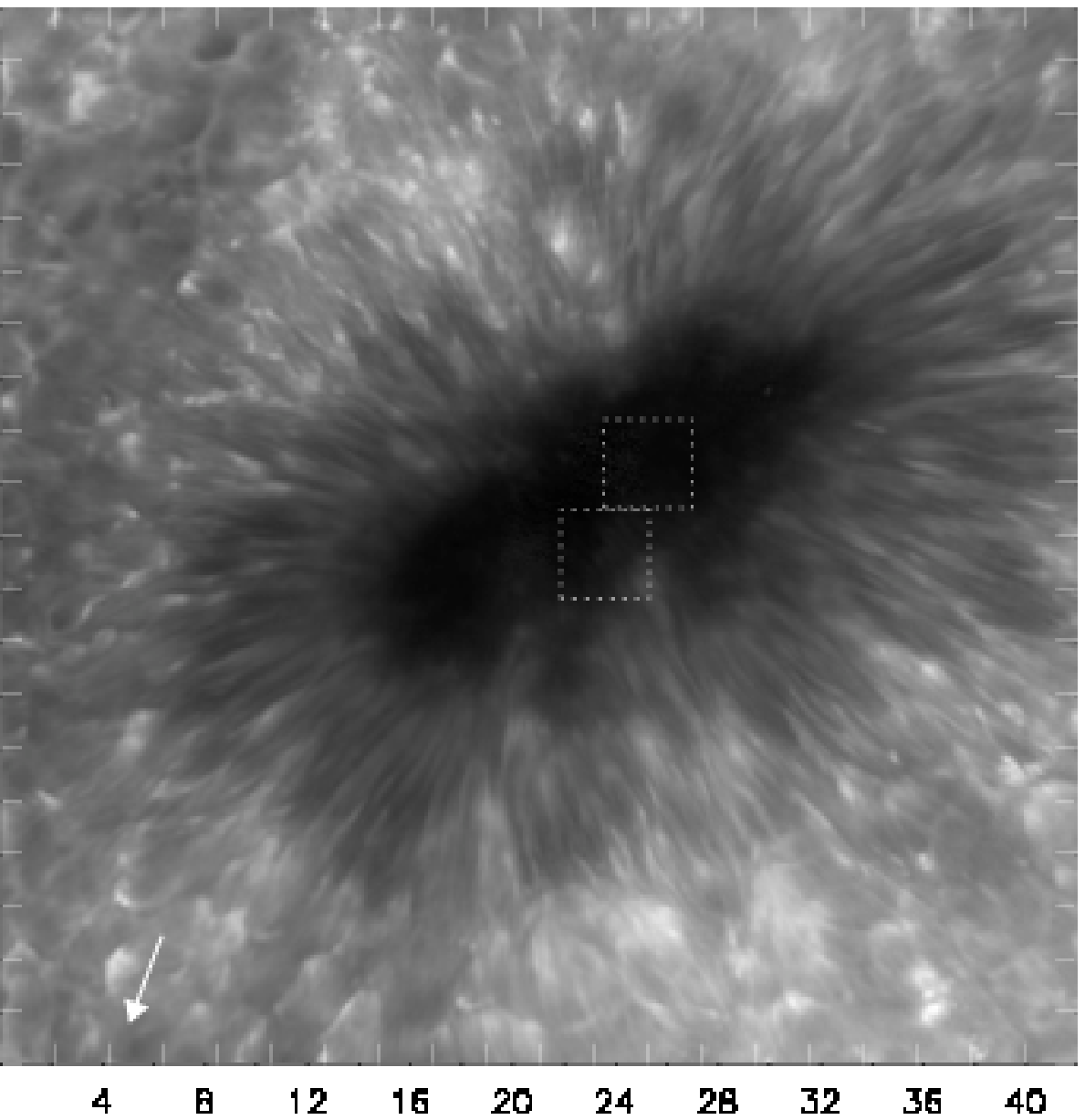}}
\resizebox{5.9cm}{!}{\includegraphics[clip=true]{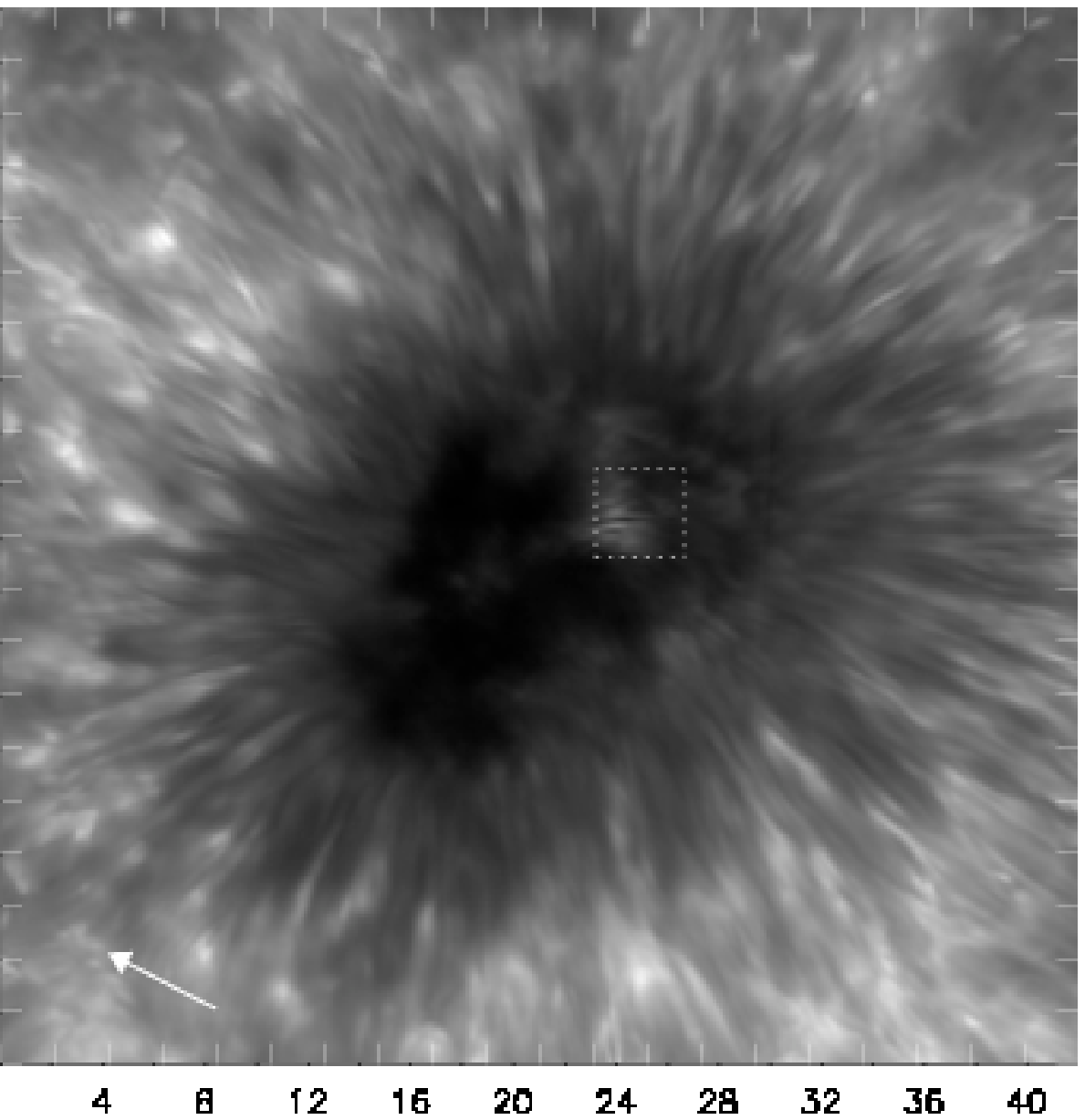}}

\caption{\footnotesize The three sunspots analysed in this work. Top row are the wideband images and bottom row the Ca II H core filter images. The arrows point towards the solar disk centre. The field of view is $42\times42$\arcsec with tickmarks at 2\arcsec intervals. The boxes delimit the regions of interest shown in Fig.~\ref{subfields}.} 
\label{fig:intro}
\end{figure*}

The observations were performed with the Swedish 1-m Solar Telescope  \citep[SST, ][]{2003SPIE.4853..341S} on three different sunspots (hereafter sunspot A, B and C). Sunspot A was associated with active region NOAA~11072 and was observed on 23~May~2010 at $\mu=0.97$. Sunspots B and C were associated with NOAA~11857, and were observed on 5~October~2013 and 7~October~2013 at $\mu=0.87$ and $\mu=0.97$, respectively. We used the standard setup to split the beam into a blue and red branch. The blue setup for both sunspots included a camera with a 10\AA\ continuum window centred at 3953.7 \AA\ (WB) and an additional camera with a 1.1\AA\ passband centred in the \ion{Ca}{ii}~H line-core (3968.4 \AA). Sunspots B and C were also observed in H$\alpha$ using the CRisp Imaging SPectropolarimeter \citep[CRISP, ][]{2006A&A...447.1111S,2008ApJ...689L..69S}. Data reduction was performed using a prototype of the code published by \cite{2014arXiv1406.0202D}. The reconstructed H$\alpha$ line-core images analysed in this work had an effective cadence of 8.4~s while for the blue data this was 13~s. The pixel scale in the blue the was 0\farcs034 per pixel corresponding to approximately 25.0~km and in the red 0\farcs071 per pixel or 52~km. All data were reconstructed with MOMFBD \citep{2005SoPh..228..191V}, correcting for 86 Karhunen-Lo\`{e}ve modes and using $128\times128$ pixel subfields for the blue data and $64\times64$ pixel subfields for the red. Adaptive optics consisting  of a tip-tilt mirror and a 37-electrode deformable mirror \citep{2003SPIE.4853..370S} were used for sunspot A while sunspots B and C were observed with a more recent 85-electrode deformable mirror setup. In Fig.~\ref{fig:intro} we show wideband and the \ion{Ca}{ii}~H images of the three sunspots for context.  

De-stretching techniques are routinely implemented in the reduction pipelines of solar data acquired from ground-based telescopes. The common procedure for the SST follows from \cite{1994ApJ...430..413S} and this is also the case in this paper. While mostly successful, the low light levels and faint structure in the umbra makes the application of de-stretching techniques challenging leading to occasional artifacts. Fortunately, these are quite straightforward to detect visually as they come in the shape of ``bubbles'' or ``wiggles''. Due to the small size of the structures analysed in this work, de-rotation, alignment for global shifts, and time-series destretching using the wideband channel as reference, were applied and then compared with frames that did not go through any resampling procedure other than MOMFBD. If the morphology of a structure or an area appeared to change in a way compatible with an artifact, then a region around that area or the whole frame was discarded. The final analysis was undertaken on the de-rotated and de-stretched frames. For all sunspots observed the seeing alternated between excellent and poor which prevented analysis of some flashes in the middle of the time sequences.


\subsection{Composite frames}
\label{flashtracking}

The umbral structures detected in Ca II H are visible during the bright component of the flashes. This component illuminates different portions of the umbra in each frame as it expands outwards in a ring pattern.  To reveal the full extent of the longest UCFs as illuminated by the flashes, a technique is used to combine frames sampling different stages of the flash ring-like expansion, into a single composite image. This technique will be referred to as ``flash tracking" and may be described as follows: each pixel in the composite image has the pixel value from the frame with the highest local spatial average in a data cube.  The spatial average is computed using a Gaussian kernel of $\sim 0.3$\arcsec FWHM as weight. Before the spatial average is computed, each frame is converted to log-scale (we note that the technique is used to study only the morphology). This technique also allows us to display UCFs from different areas of the umbra in the same image.  It is used in figures \ref{trails} and \ref{spotc}, and then only in the panels noted. Time windows of 104 seconds were used for the composite frames.


\section{Results and Discussion}
\label{results}

\begin{figure*}
  \centering
  \def\lab#1{\begin{minipage}[b]{4mm}
      #1\vspace{12mm}
    \end{minipage}}
  \def\tile#1{\resizebox{!}{2.5cm}{ \includegraphics{#1}}}
\begin{tabular}{c@{\hspace{0.1mm}}c@{\hspace{0.1mm}}c@{\hspace{0.1mm}}c@{\hspace{0.1mm}}c@{\hspace{0.1mm}}c@{\hspace{0.1mm}}c@{\hspace{0.1mm}}c}

      \hspace{-2mm} Spot &  t=0 & t=1.5  & t=6.1 &  t=10.6 & t=16.7   \\
       \lab{A}  &   \resizebox{!}{2.5cm}{ \includegraphics{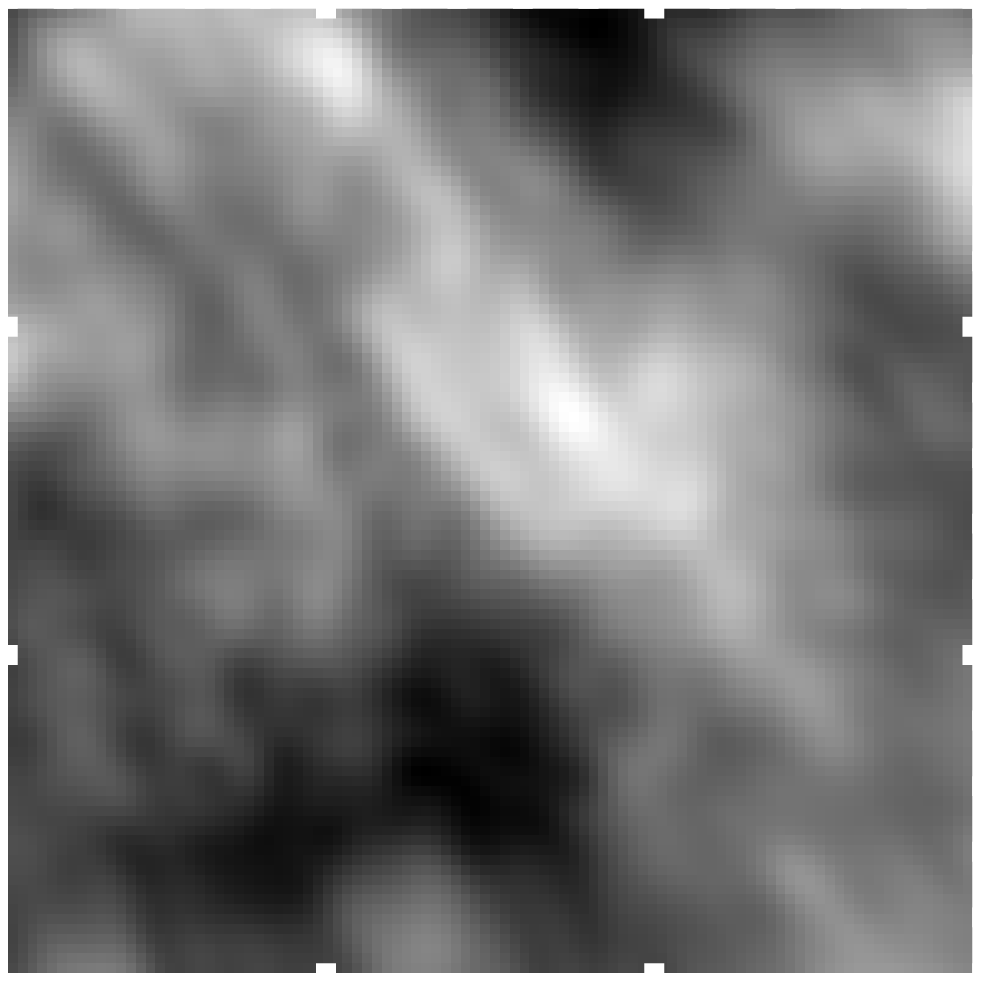}} &   \resizebox{!}{2.5cm}{ \includegraphics{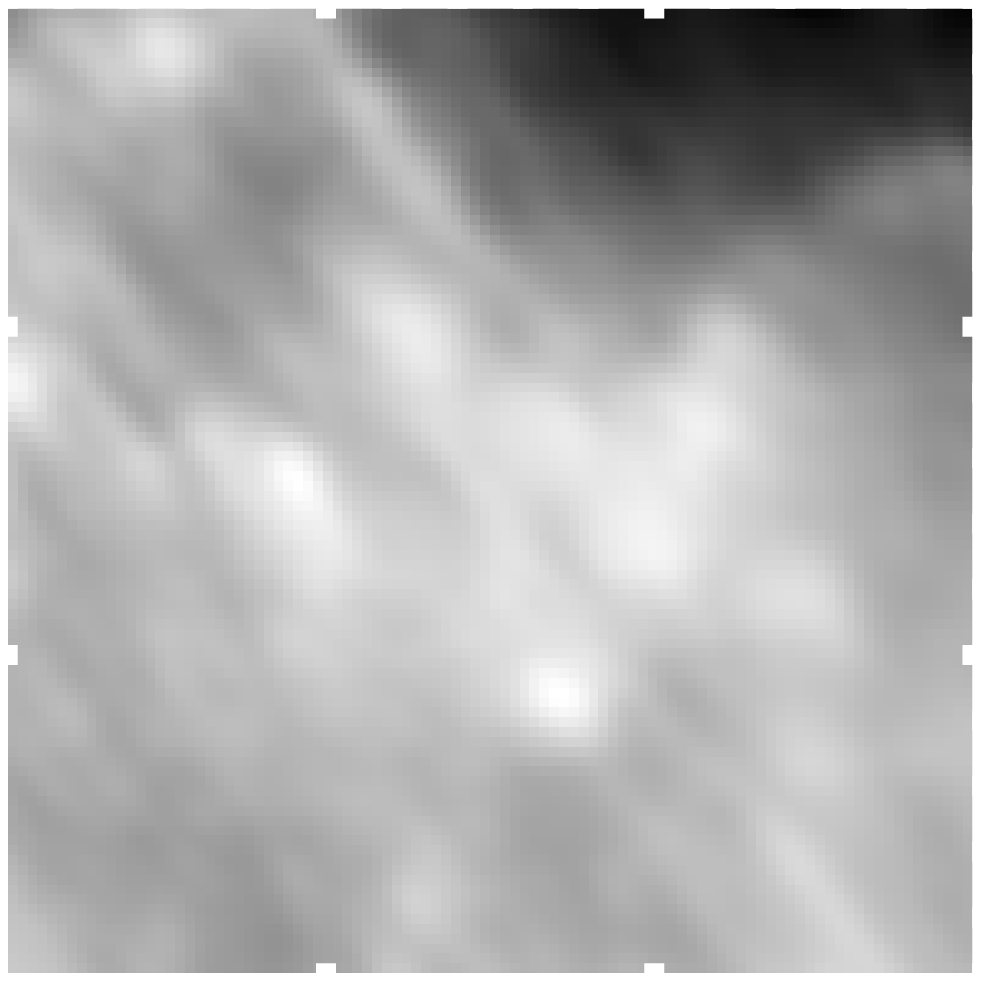}} &  \resizebox{!}{2.5cm}{ \includegraphics{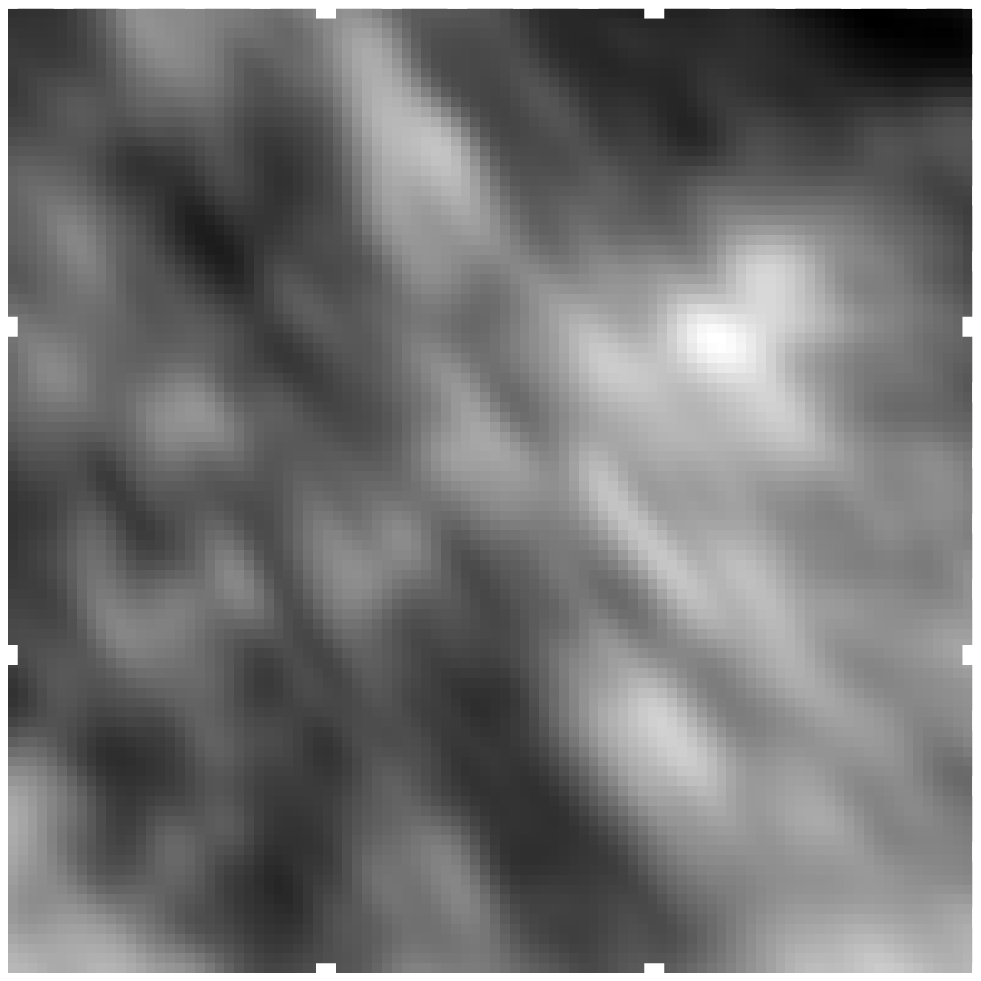}} &  \resizebox{!}{2.5cm}{ \includegraphics{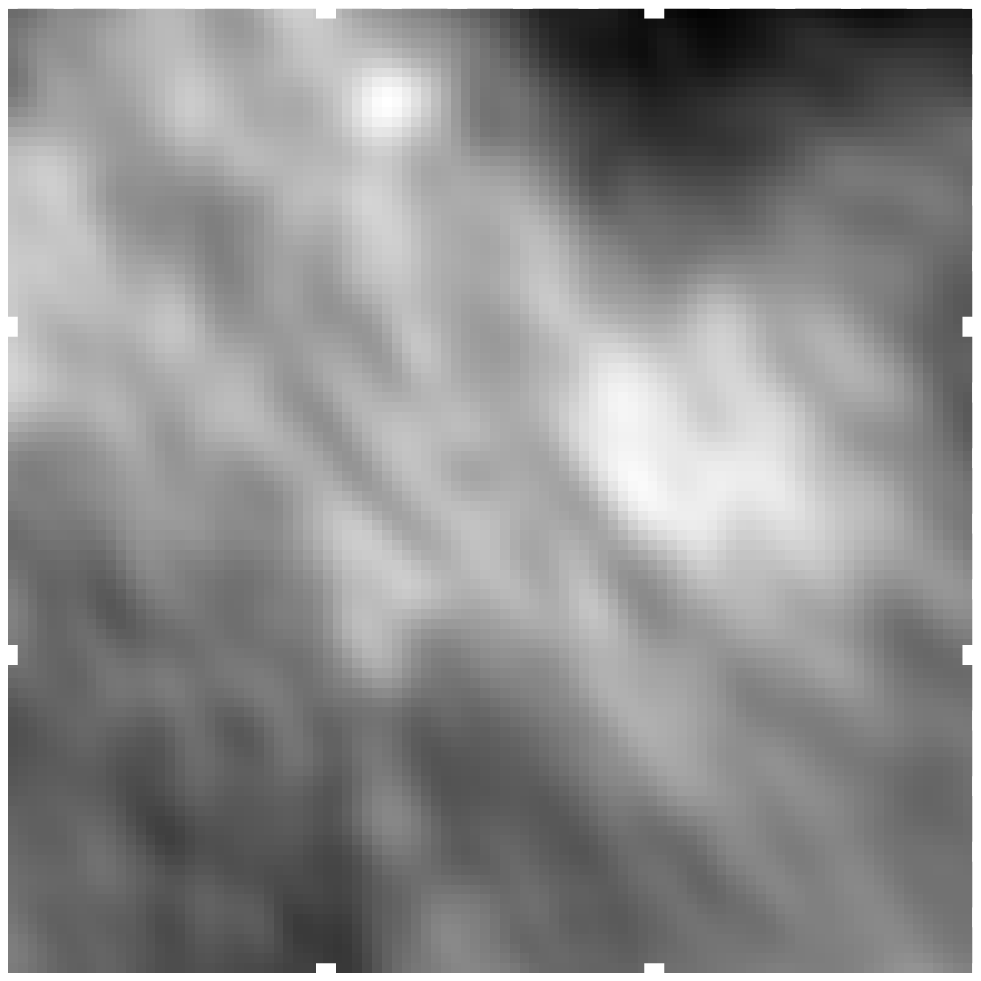}} &  \resizebox{!}{2.5cm}{ \includegraphics{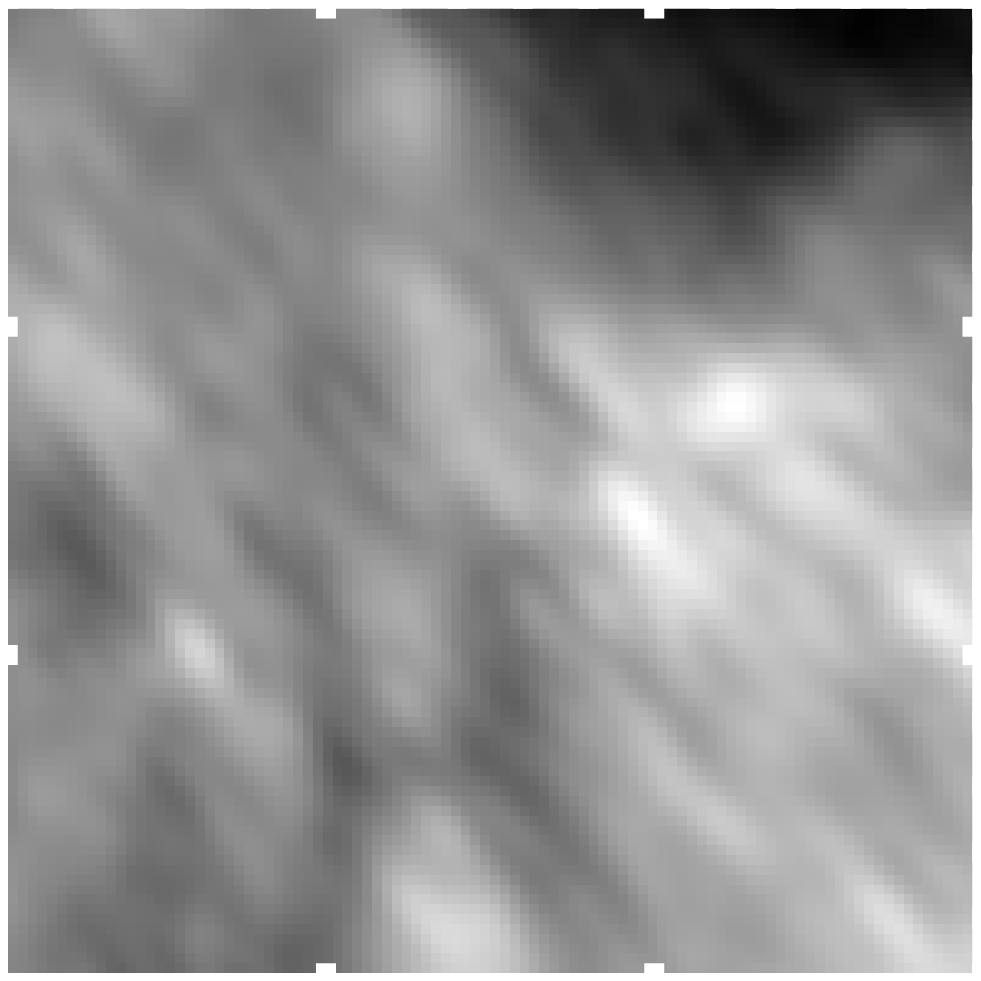}}  \\

 &  t=0.14 & t=1.40  &  t=3.08 & t=5.18 &  t=7.28  \\  
       \lab{B1}  &   \resizebox{!}{2.5cm}{ \includegraphics{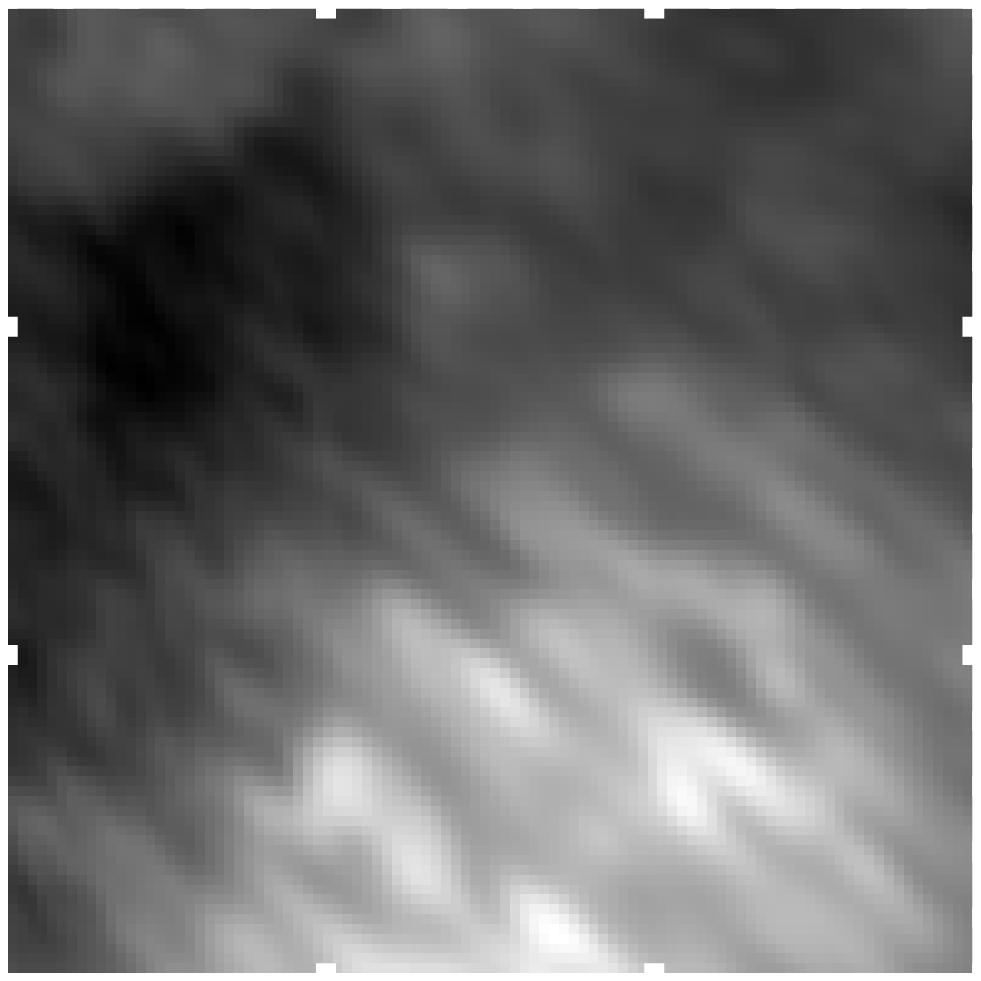}} &  \resizebox{2.5cm}{!}{\includegraphics[clip=true]{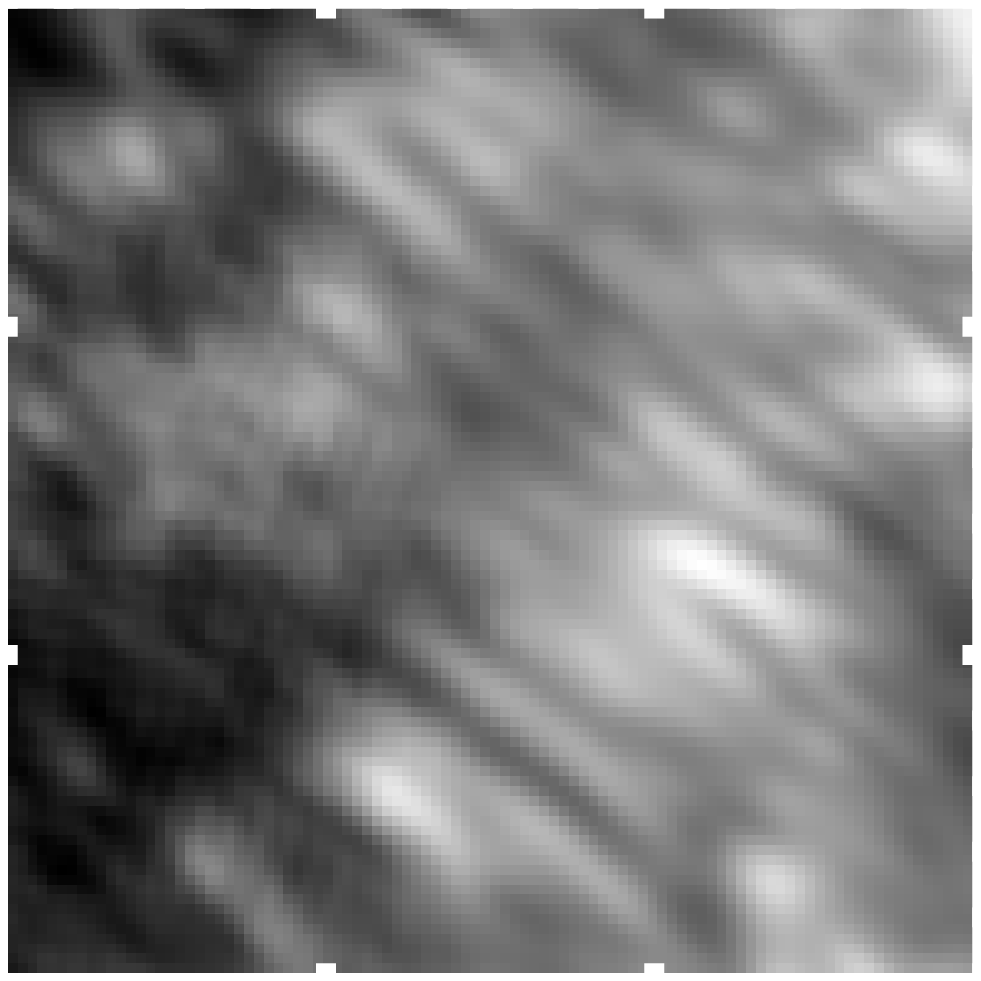}} & \resizebox{2.5cm}{!}{\includegraphics[clip=true]{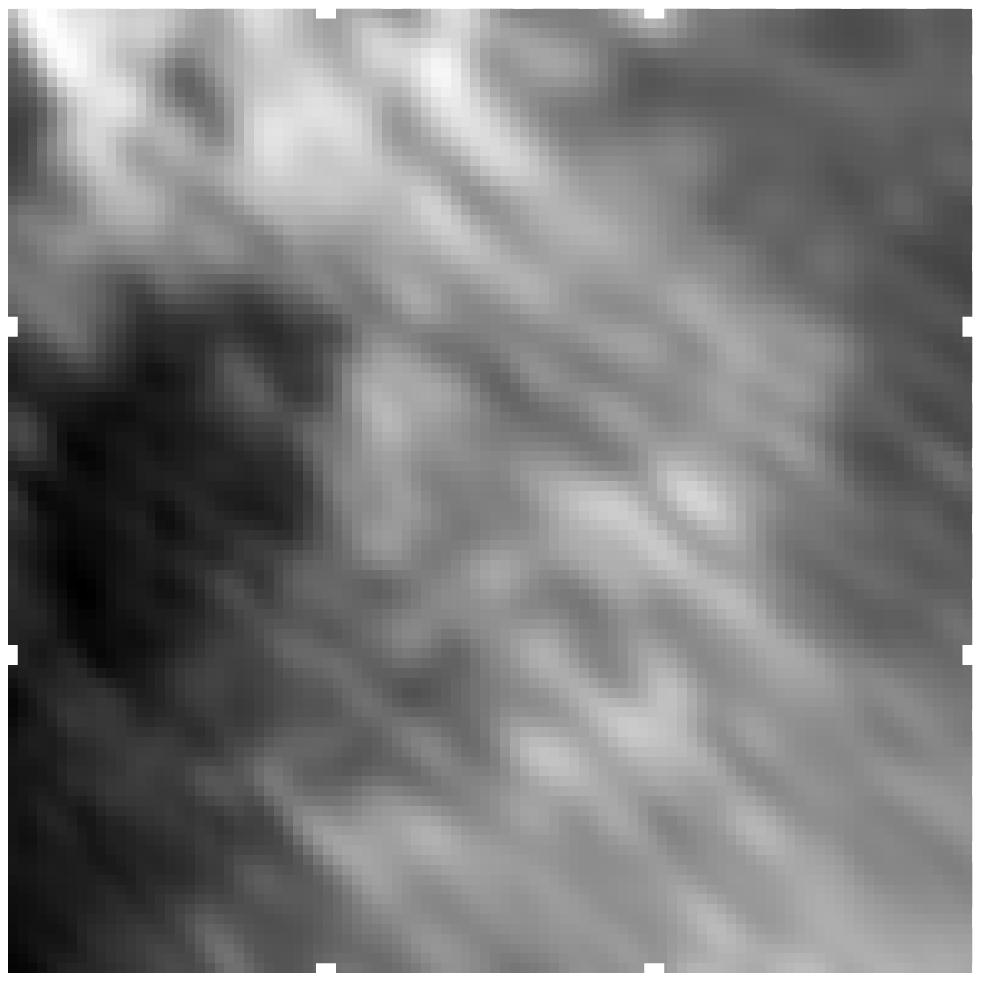}} & \resizebox{2.5cm}{!}{\includegraphics[clip=true]{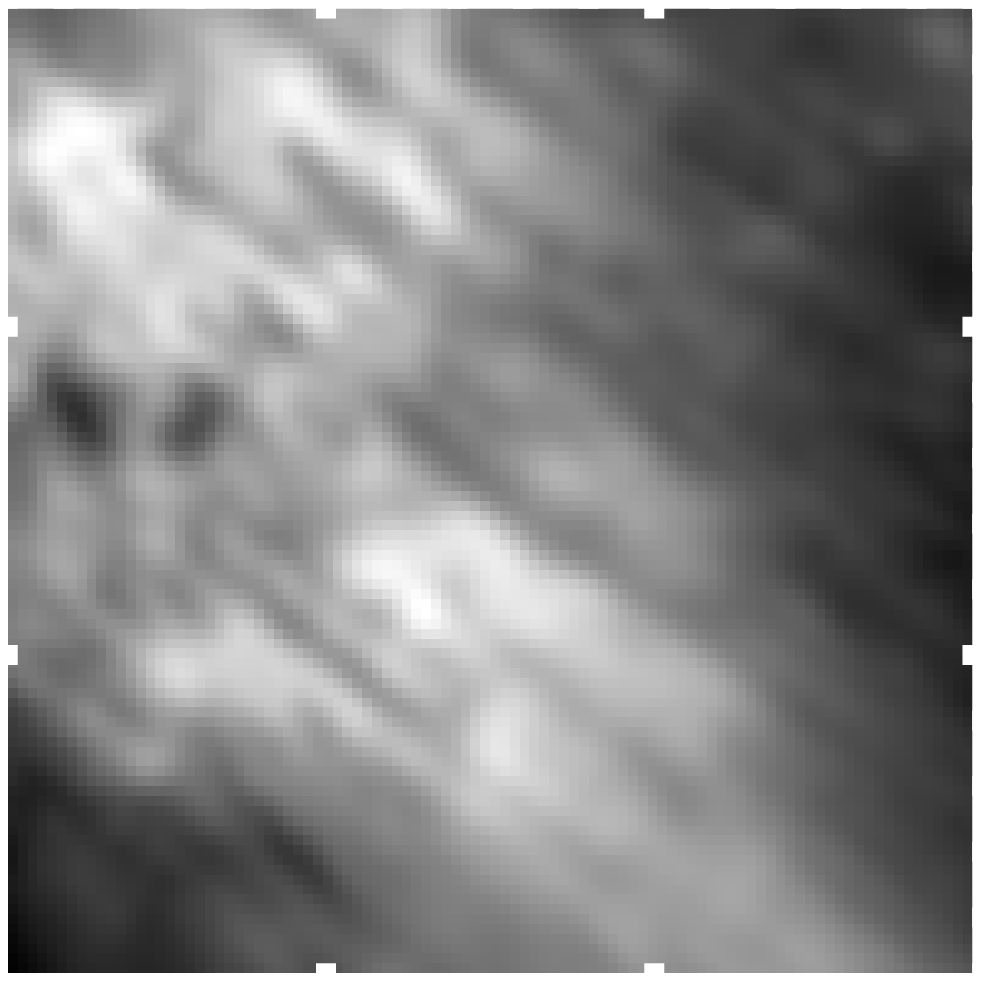}} & \resizebox{2.5cm}{!}{\includegraphics[clip=true]{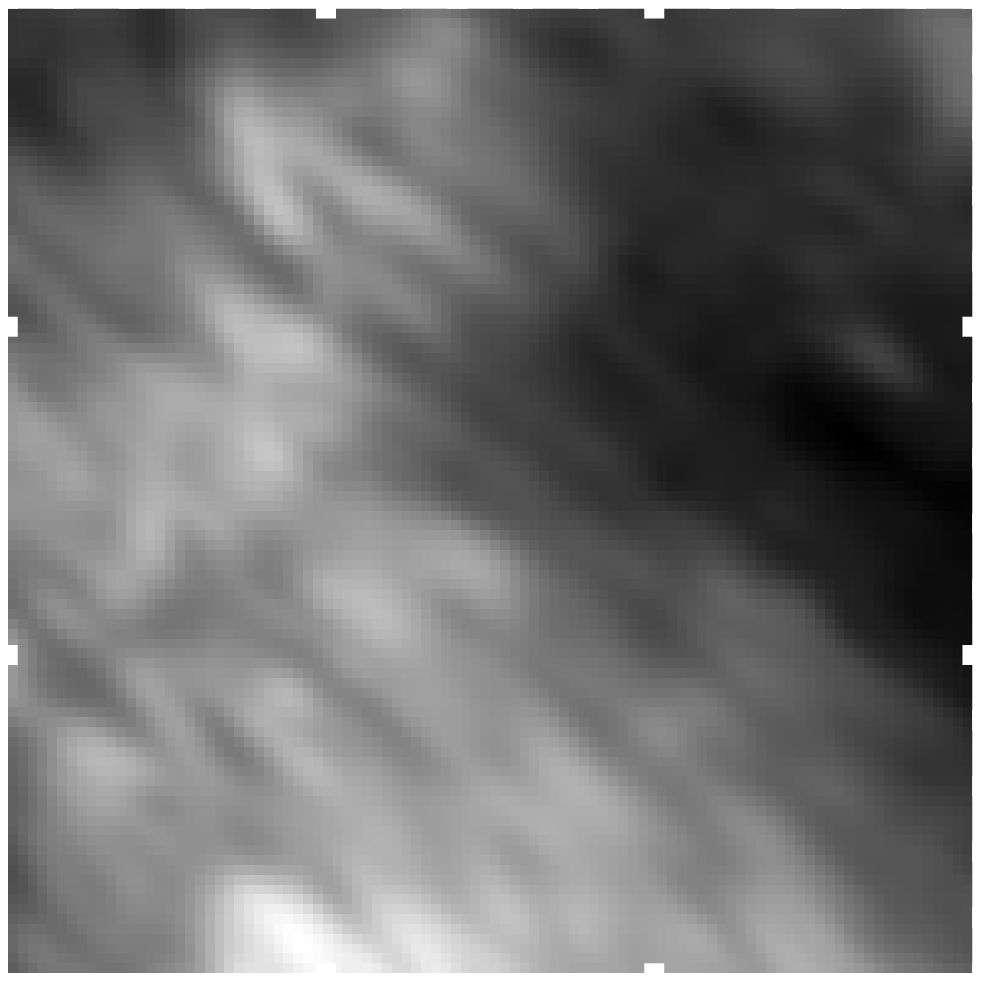}}  \\

      &  t=0.42 & t=1.4  & t=4.20 &  t=6.86 & t=8.40 & t=12.46  \\
   \lab{B2}  &  \resizebox{2.5cm}{!}{\includegraphics[clip=true]{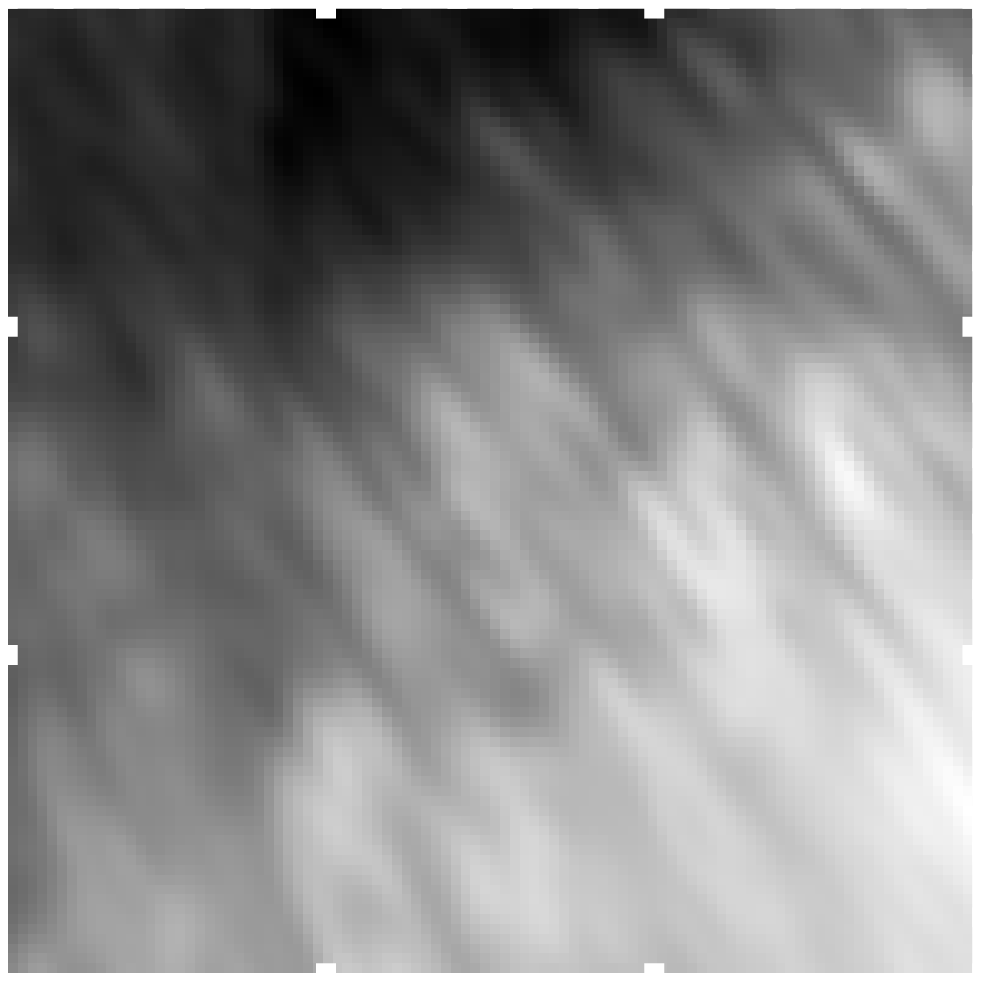}} & \resizebox{2.5cm}{!}{\includegraphics[clip=true]{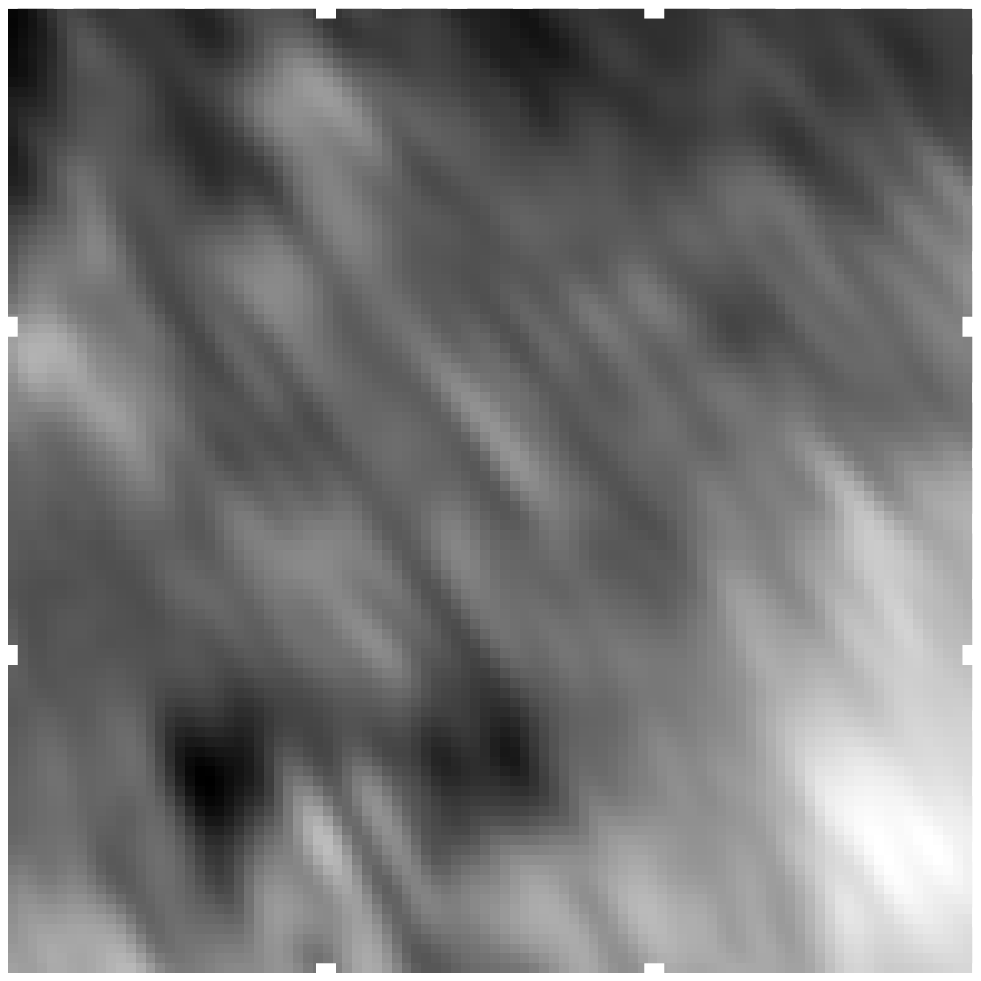}} & \resizebox{2.5cm}{!}{\includegraphics[clip=true]{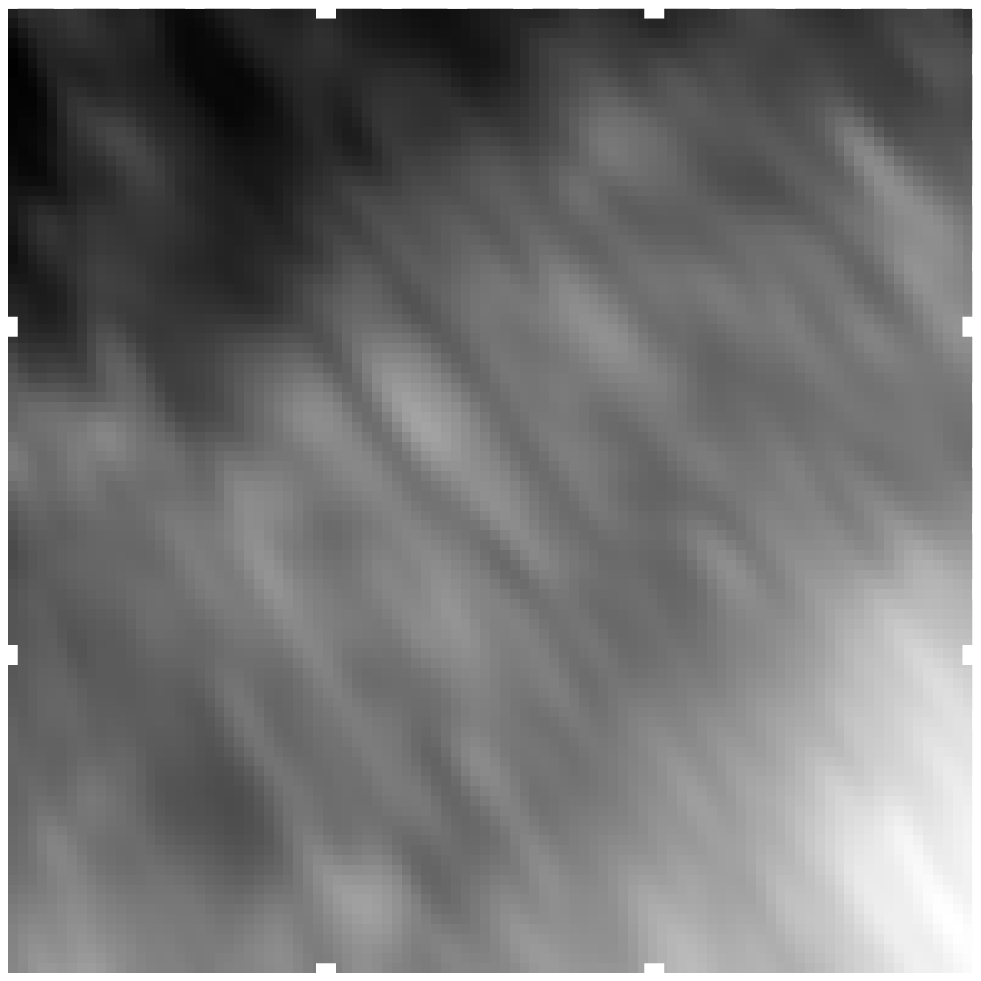}} & \resizebox{2.5cm}{!}{\includegraphics[clip=true]{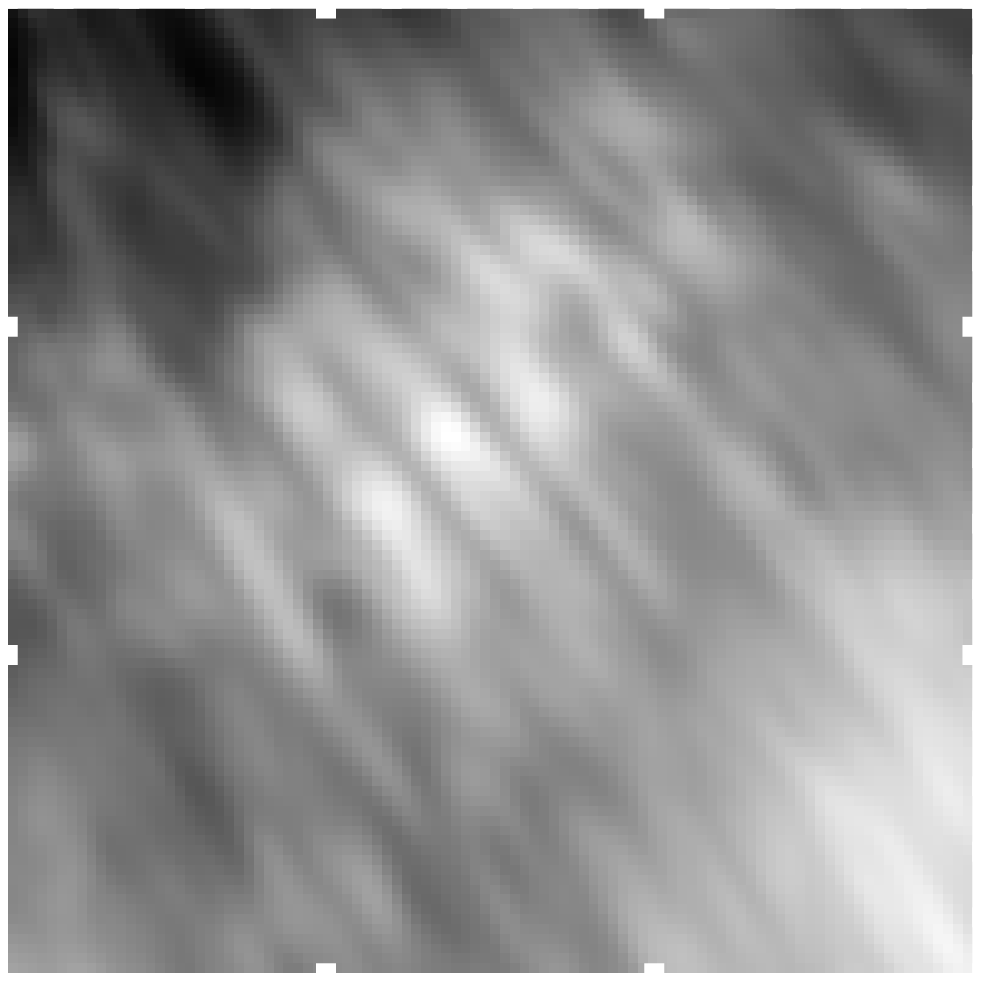}} & \resizebox{2.5cm}{!}{\includegraphics[clip=true]{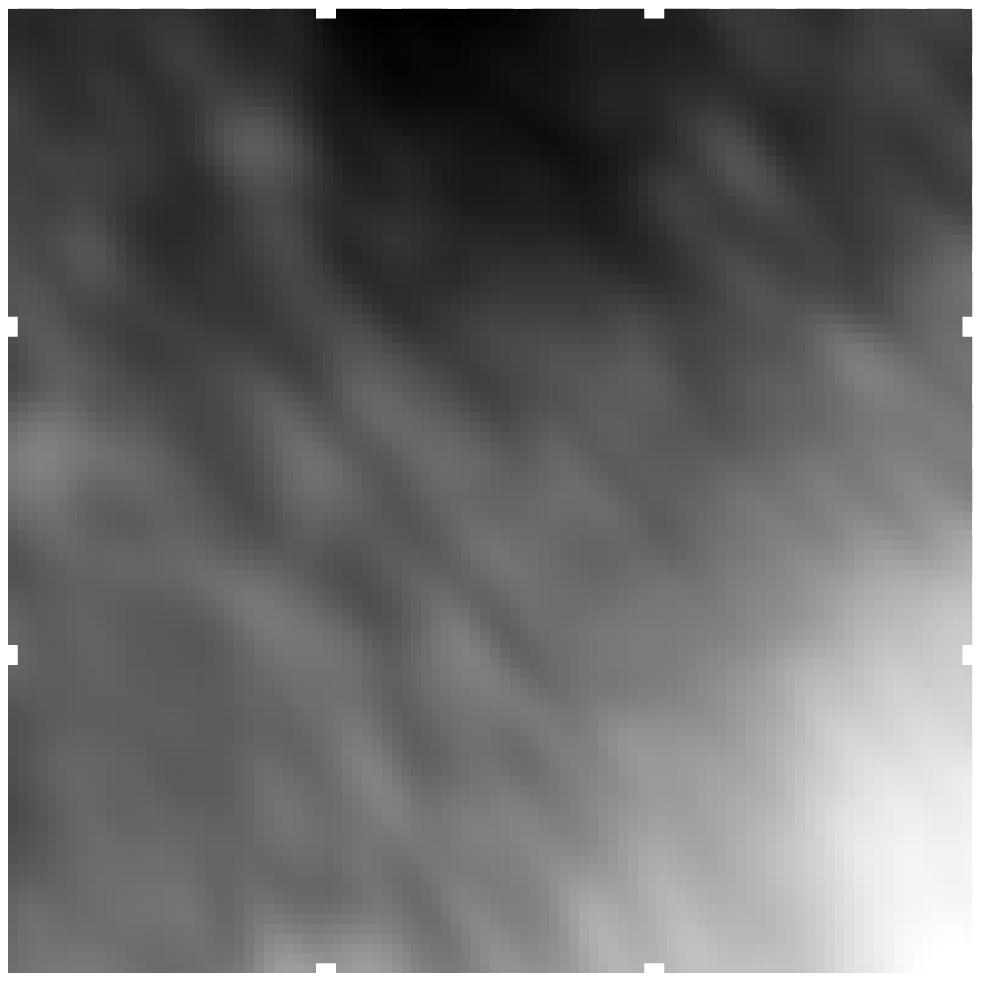}} & \resizebox{2.5cm}{!}{\includegraphics[clip=true]{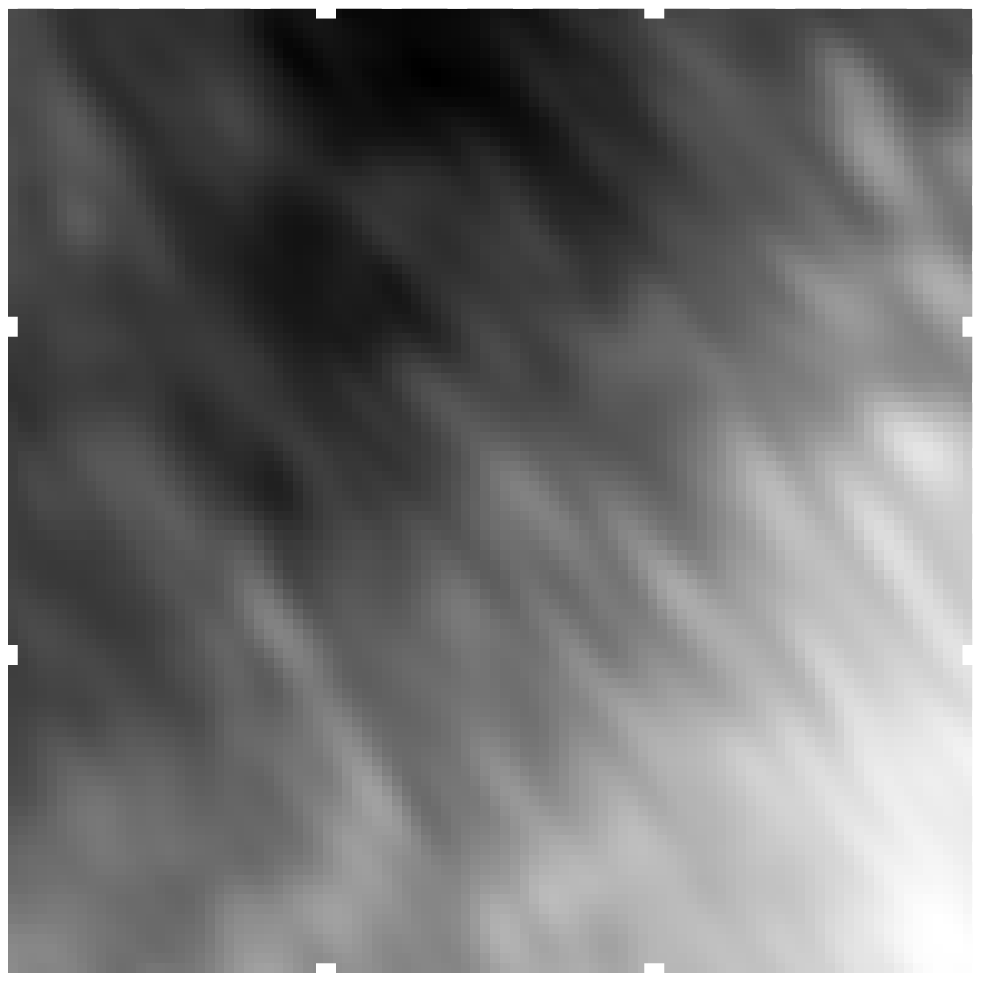}} \\

      &  t=1.26 & t=6.86  & t=9.66 &  t=11.90 & t=14.42 &  t=17.50    \\
 \lab{C}  & \resizebox{2.5cm}{!}{\includegraphics[clip=true]{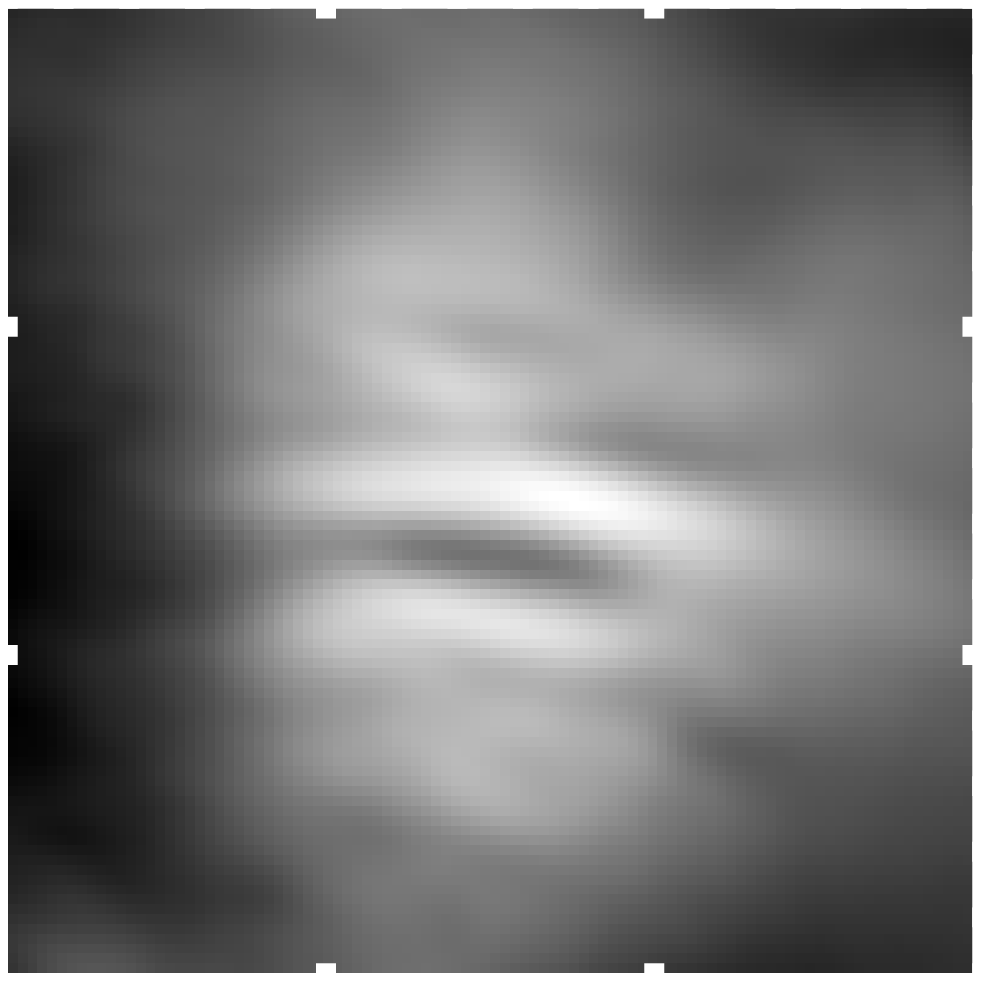}} & \resizebox{2.5cm}{!}{\includegraphics[clip=true]{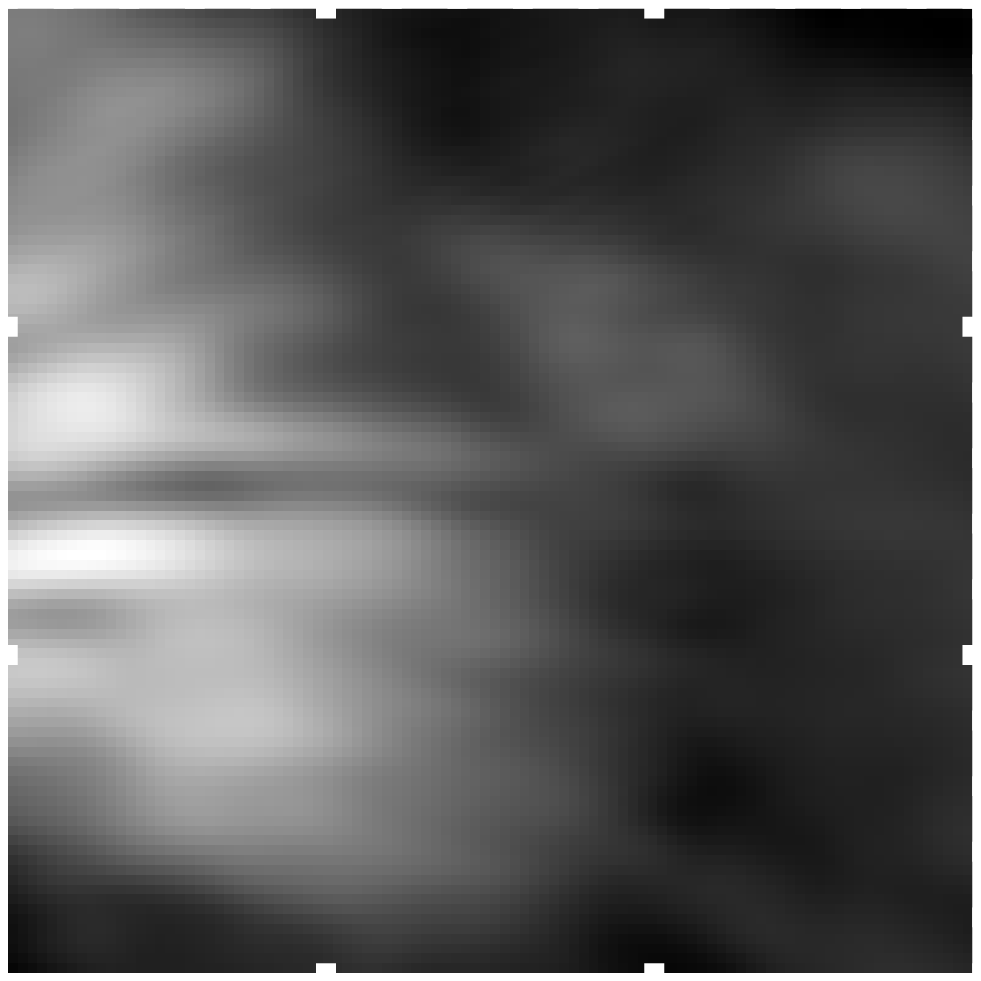}} & \resizebox{2.5cm}{!}{\includegraphics[clip=true]{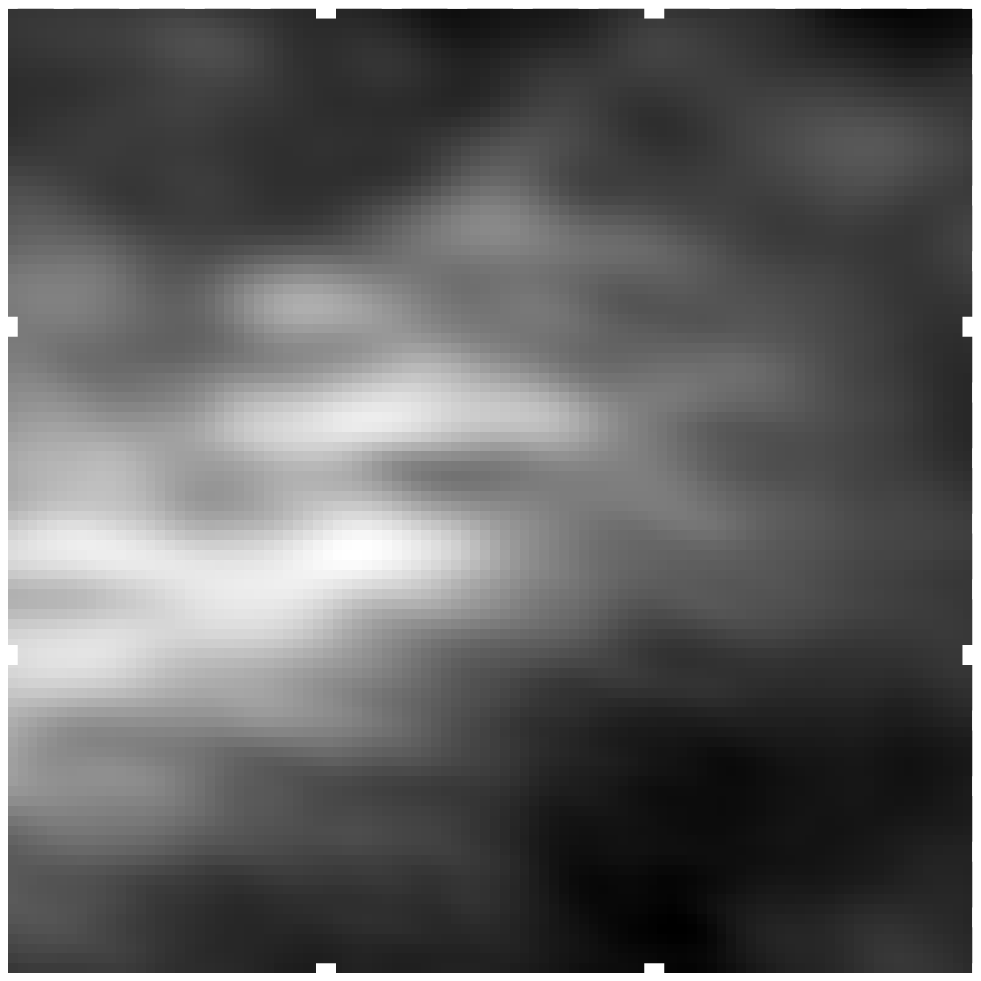}} & \resizebox{2.5cm}{!}{\includegraphics[clip=true]{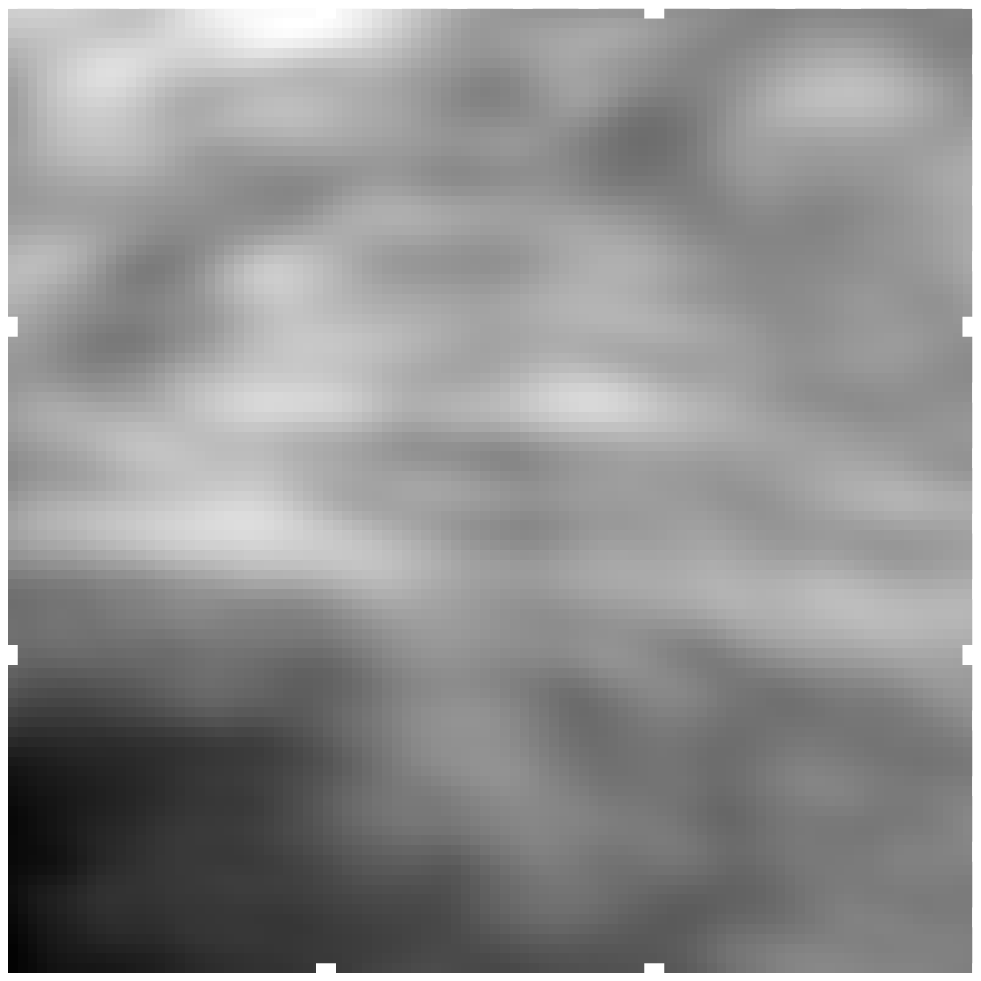}} & \resizebox{2.5cm}{!}{\includegraphics[clip=true]{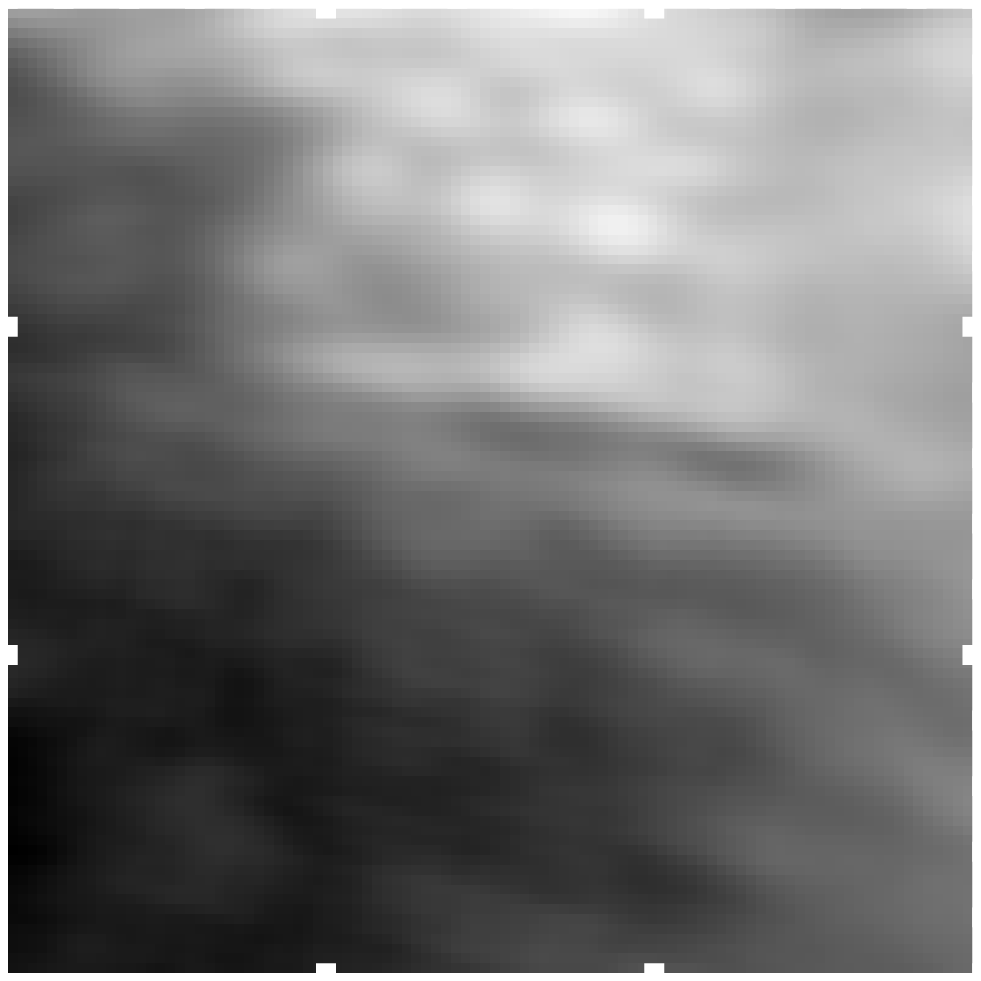}} & \resizebox{2.5cm}{!}{\includegraphics[clip=true]{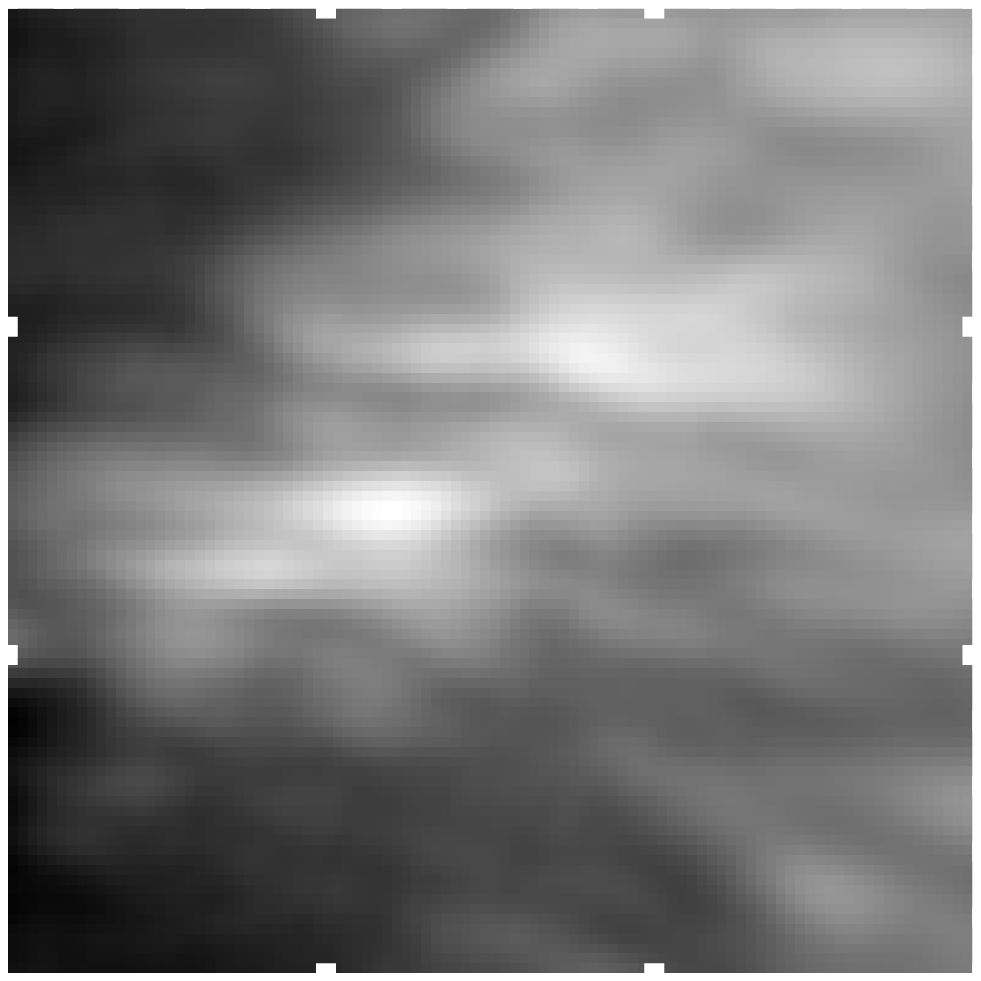}}

    \end{tabular}
  \caption{\footnotesize Selected umbra subfields of the sunspots under investigation. Logarithmic scale was used. The times for sunspot~B and C were matched with those of the respective time series (online material of Fig.~\ref{trails} and Fig.~\ref{spotc}). Major tickmarks at 1\arcsec intervals, t is in minutes.}  
      \label{subfields}
\end{figure*}

\begin{figure*}[!htb]
  \def\lab#1{\begin{minipage}[b]{32mm}#1
  \vspace{1mm}
    \end{minipage}}
  \centering

 \lab{\textbf{t = 4.90 min}}  
 \lab{\hspace{14mm} \textbf{t = 7.00 min}}  

\resizebox{7.3cm}{!}{\includegraphics[clip=true]{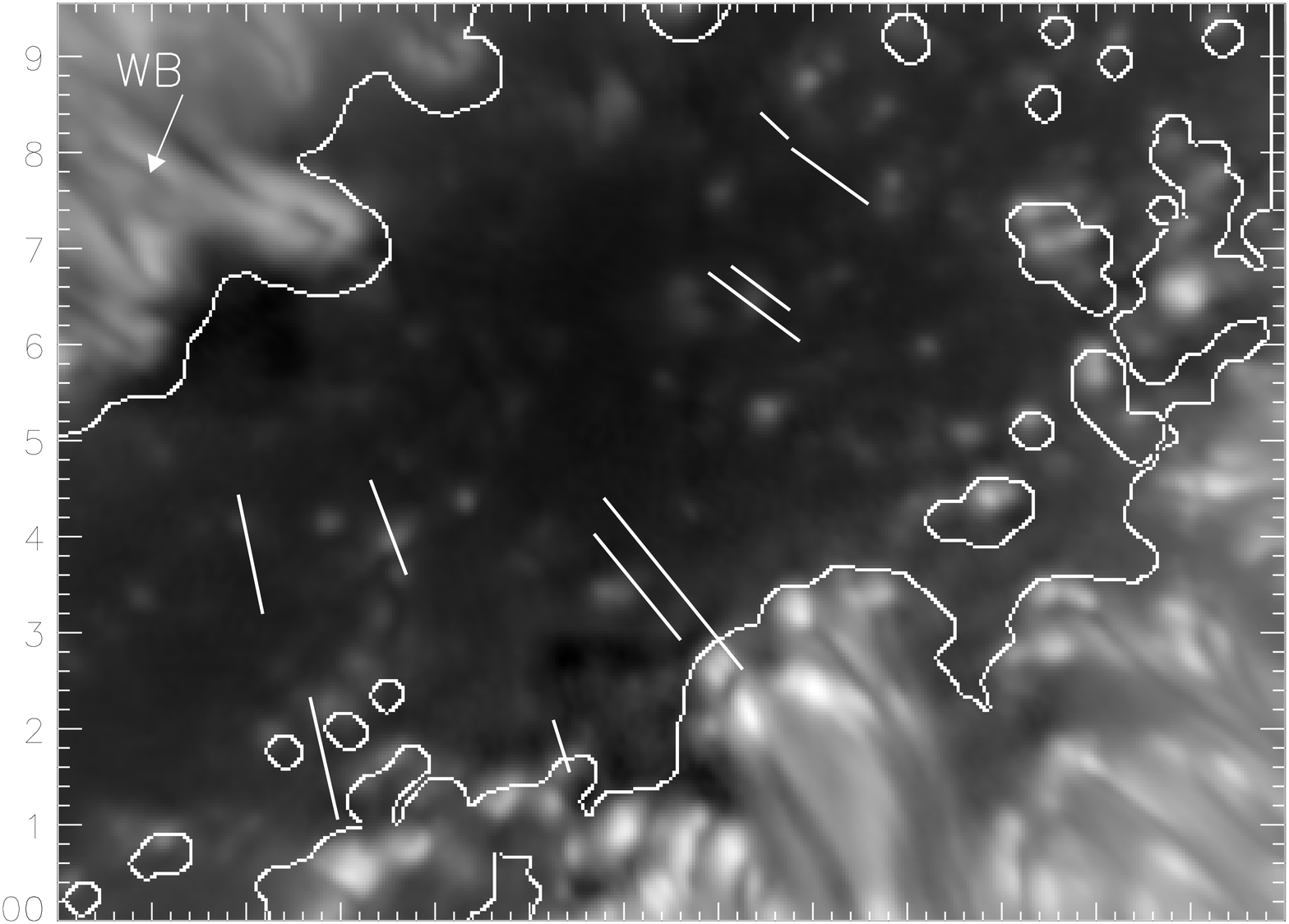}}
\resizebox{7.3cm}{!}{\includegraphics[clip=true]{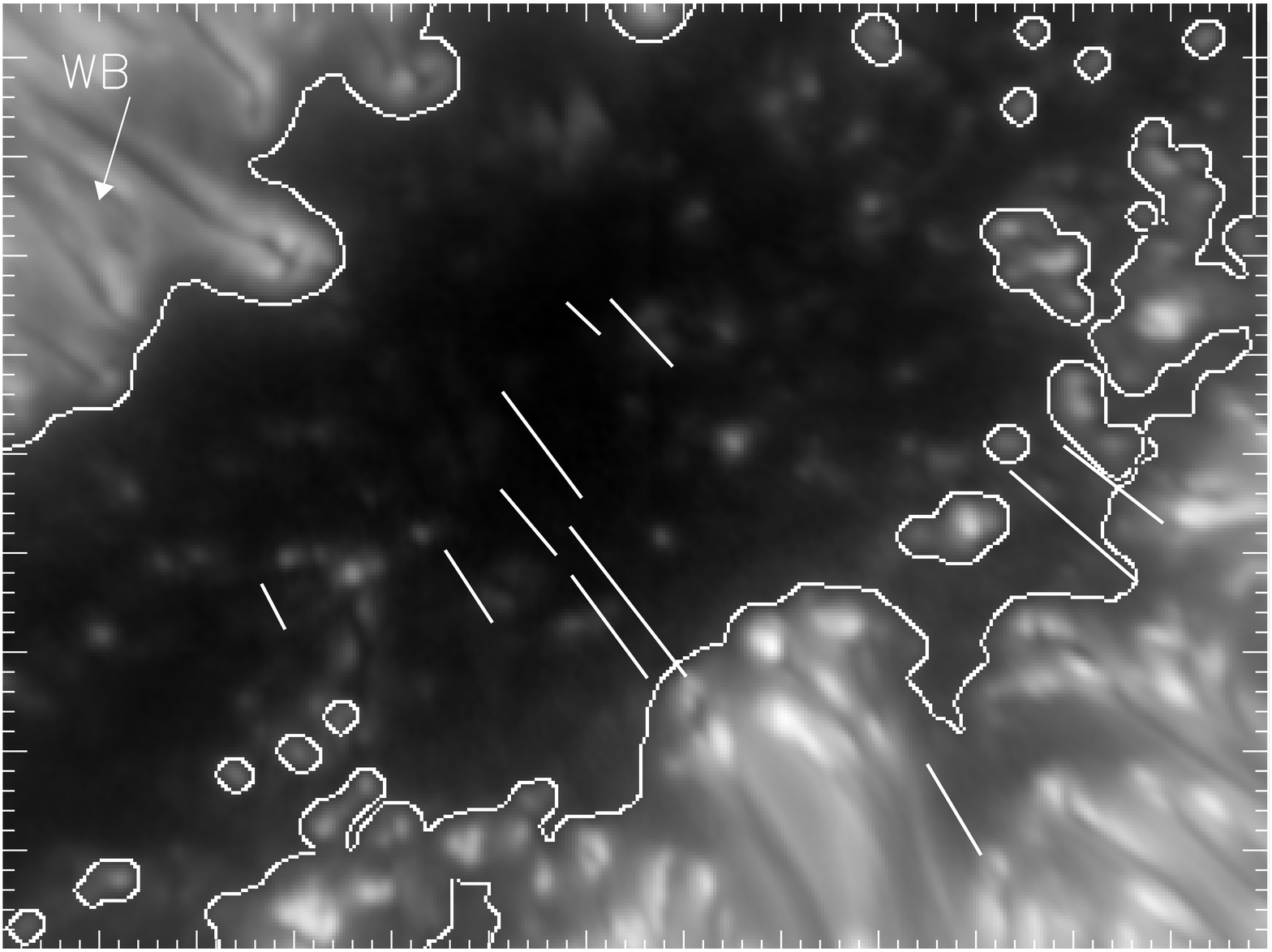}}

\resizebox{7.3cm}{!}{\includegraphics[clip=true]{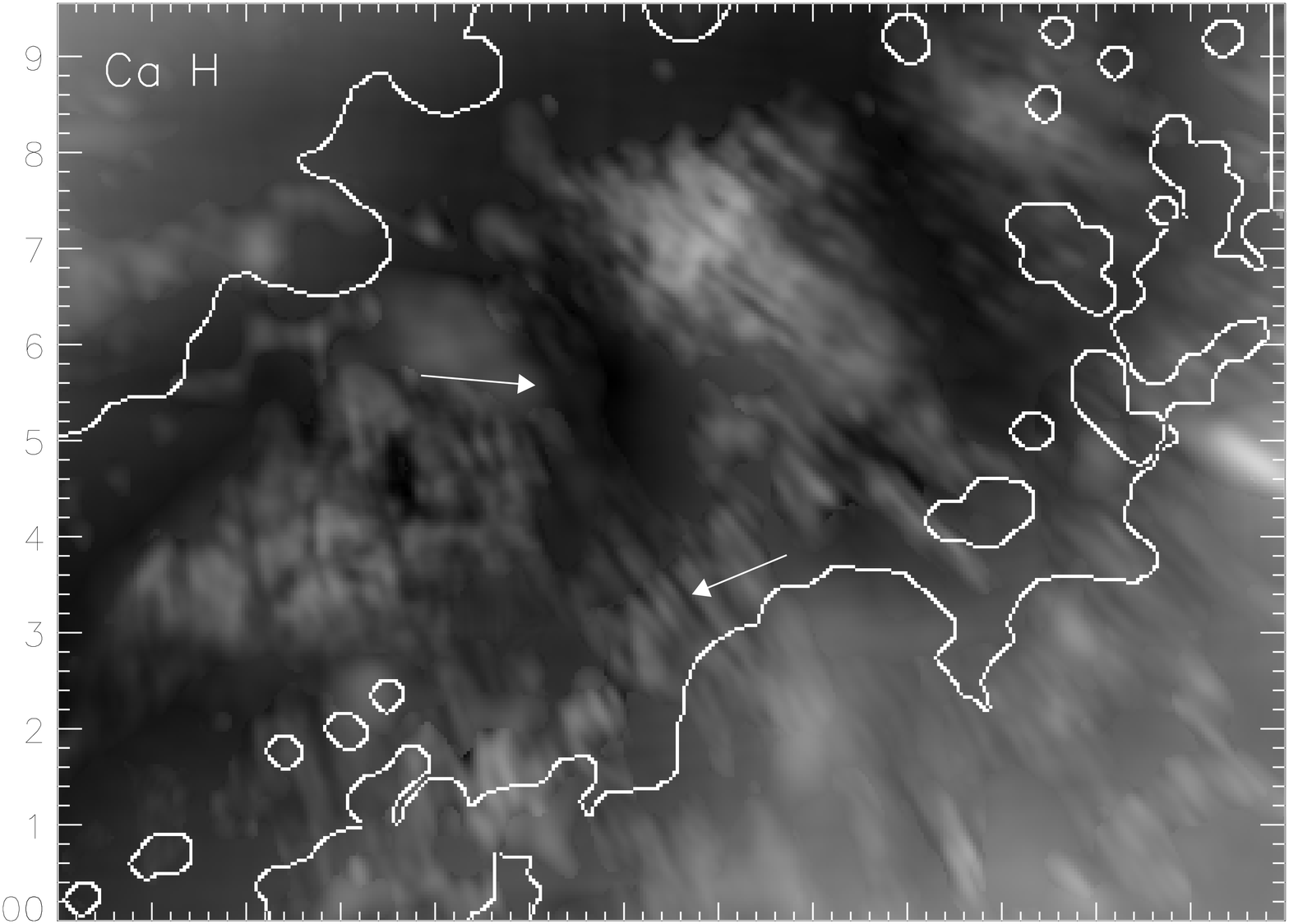}}
\resizebox{7.3cm}{!}{\includegraphics[clip=true]{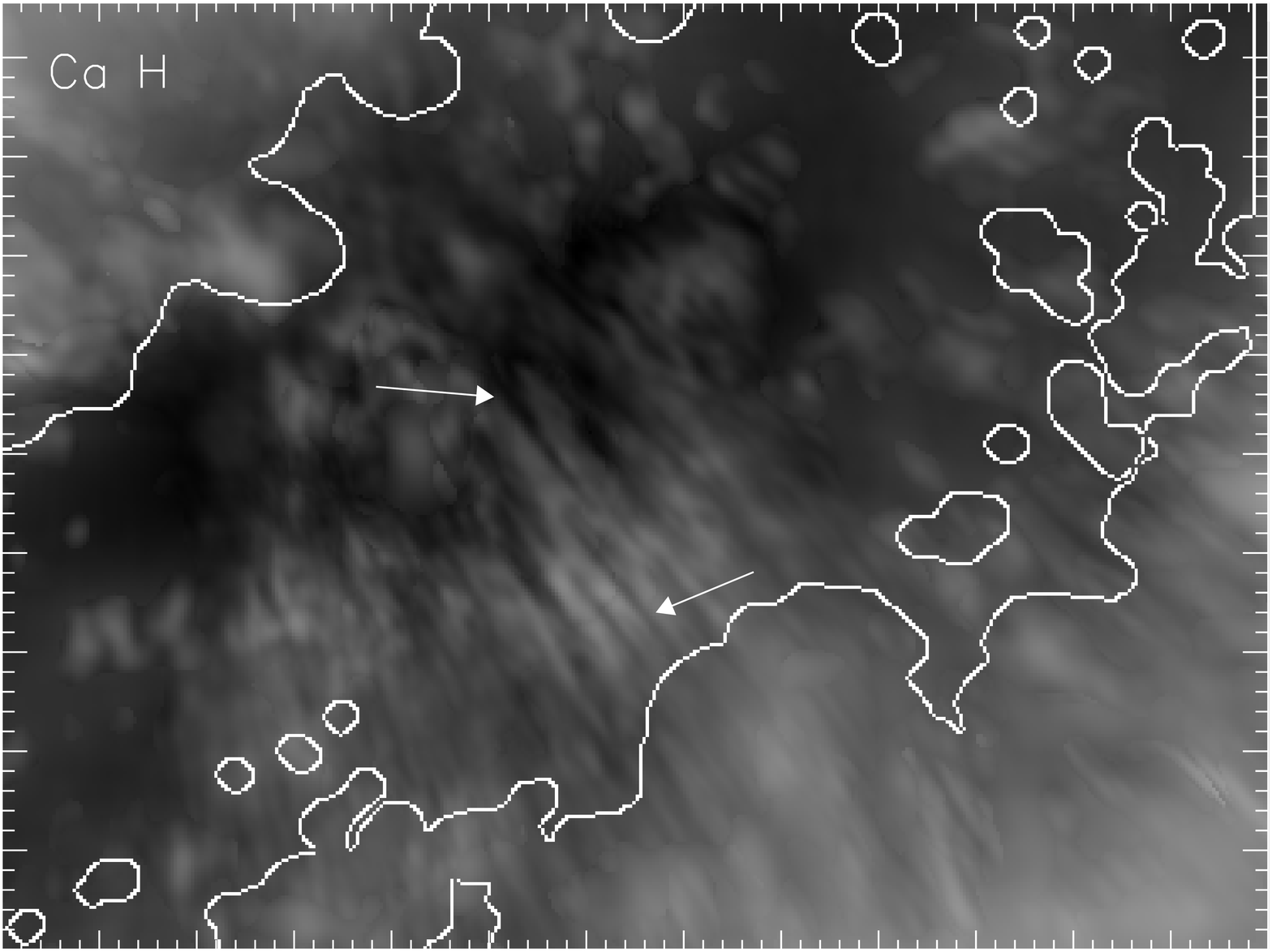}}

\resizebox{7.3cm}{!}{\includegraphics[clip=true]{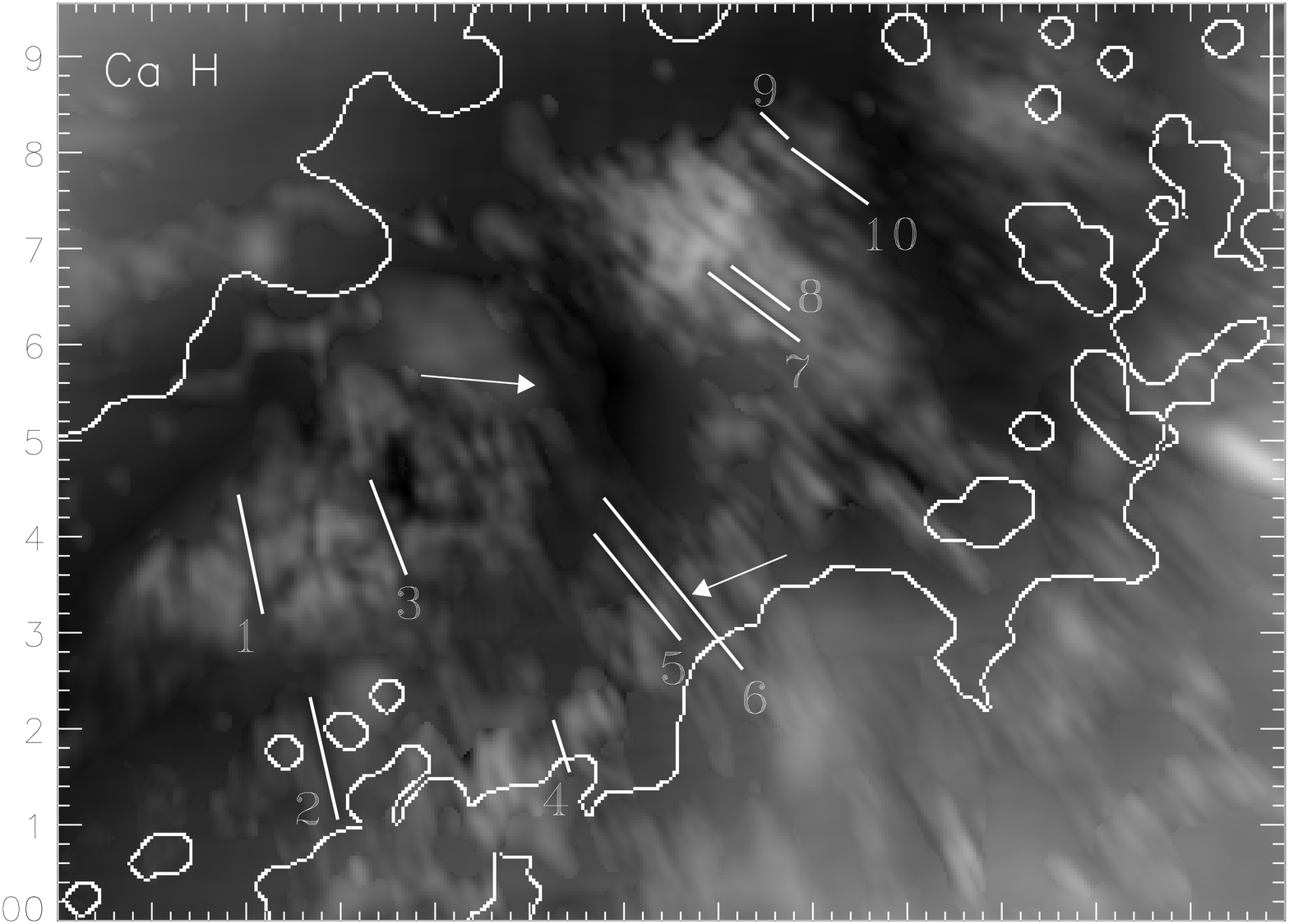}}
\resizebox{7.3cm}{!}{\includegraphics[clip=true]{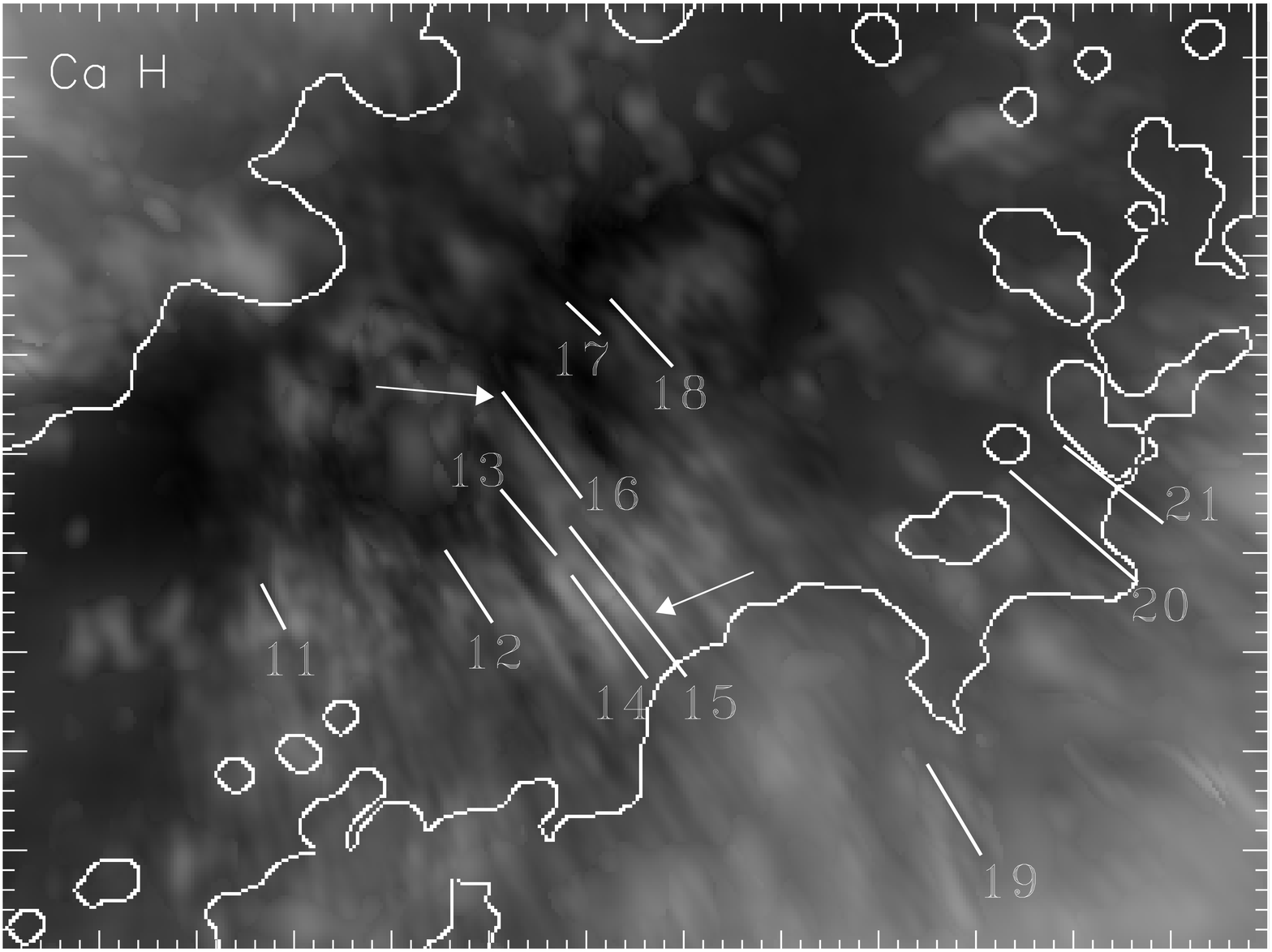}}

\resizebox{7.3cm}{!}{\includegraphics[clip=true]{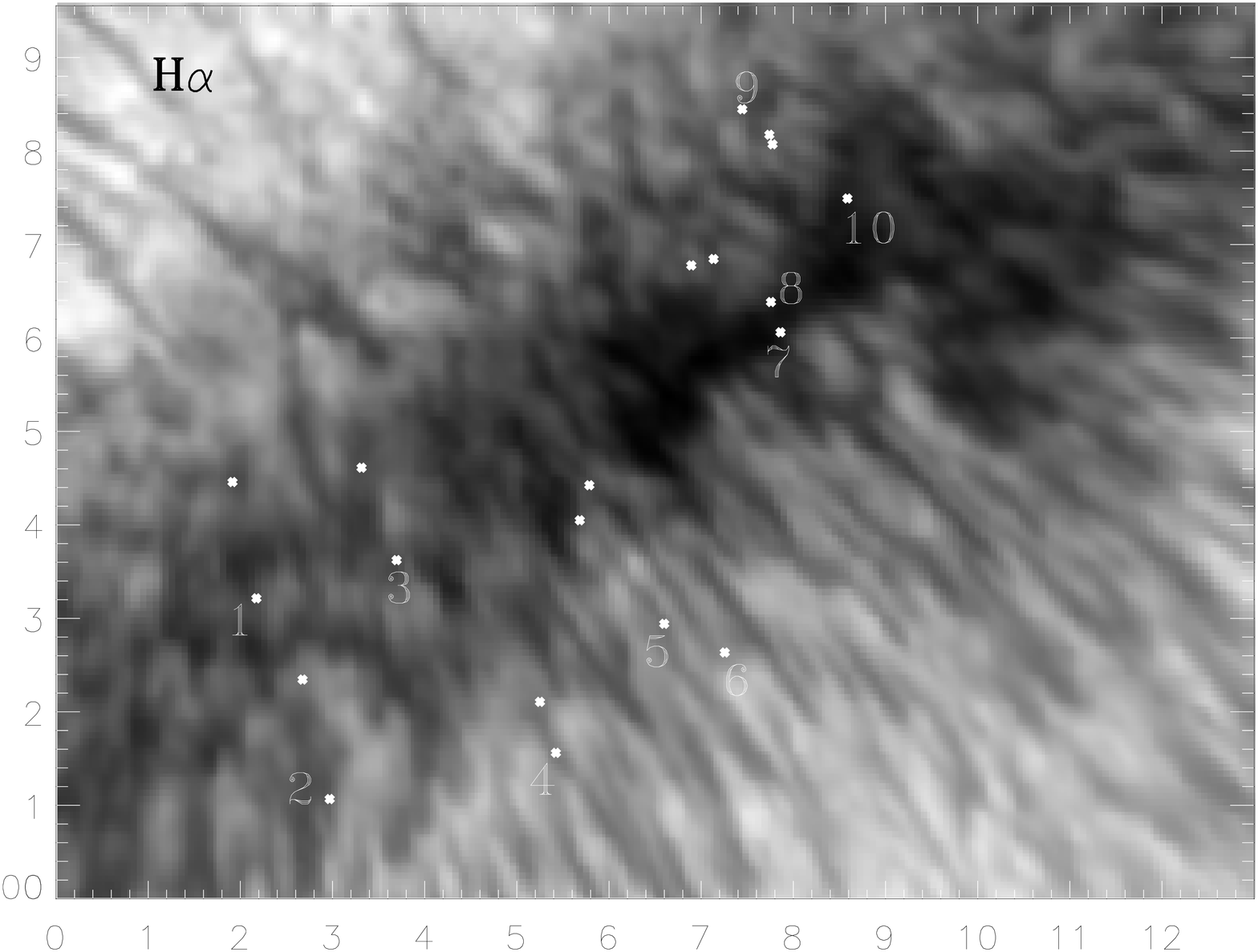}}
\resizebox{7.3cm}{!}{\includegraphics[clip=true]{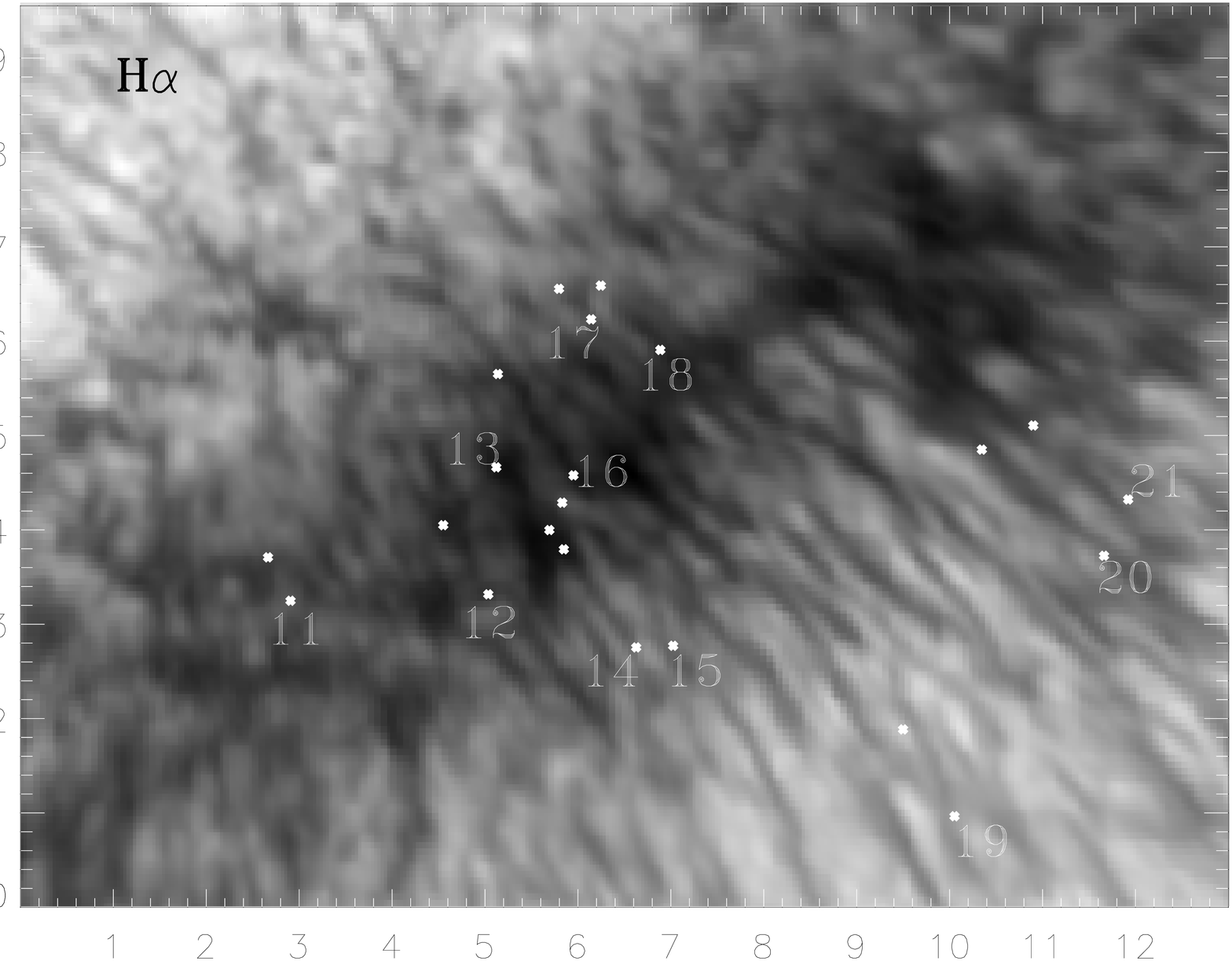}}

\caption{\footnotesize The two columns show two consecutive flashes, in differen wavelengths, that have occurred 2.10 minutes apart in Spot B. Top row: the WB images. The top left arrow indicates the disk-centre direction. Second row: Ca II H core images composed using the technique from Section~\ref{flashtracking}. The arrow on the right points at a pair of two dark structures visible in both columns. The arrow on the left points at the base of a triangular shaped structure. Third row: same as the second row but with the traces used to measure lengths overplotted. The contours show the umbra outlined using intensity thresholds in the WB. Bottom row: corresponding frames observed in H-alpha with white markers showing the extremities of the traces overplotted in the third row. Major tickmarks every 1\arcsec. The \ion{Ca}{ii}~H core frames, without composition, are shown in the movie provided with the online version.}
\label{trails}
\end{figure*}

\begin{figure*}[!htb]
  \centering

\resizebox{7.5cm}{!}{\includegraphics[clip=true]{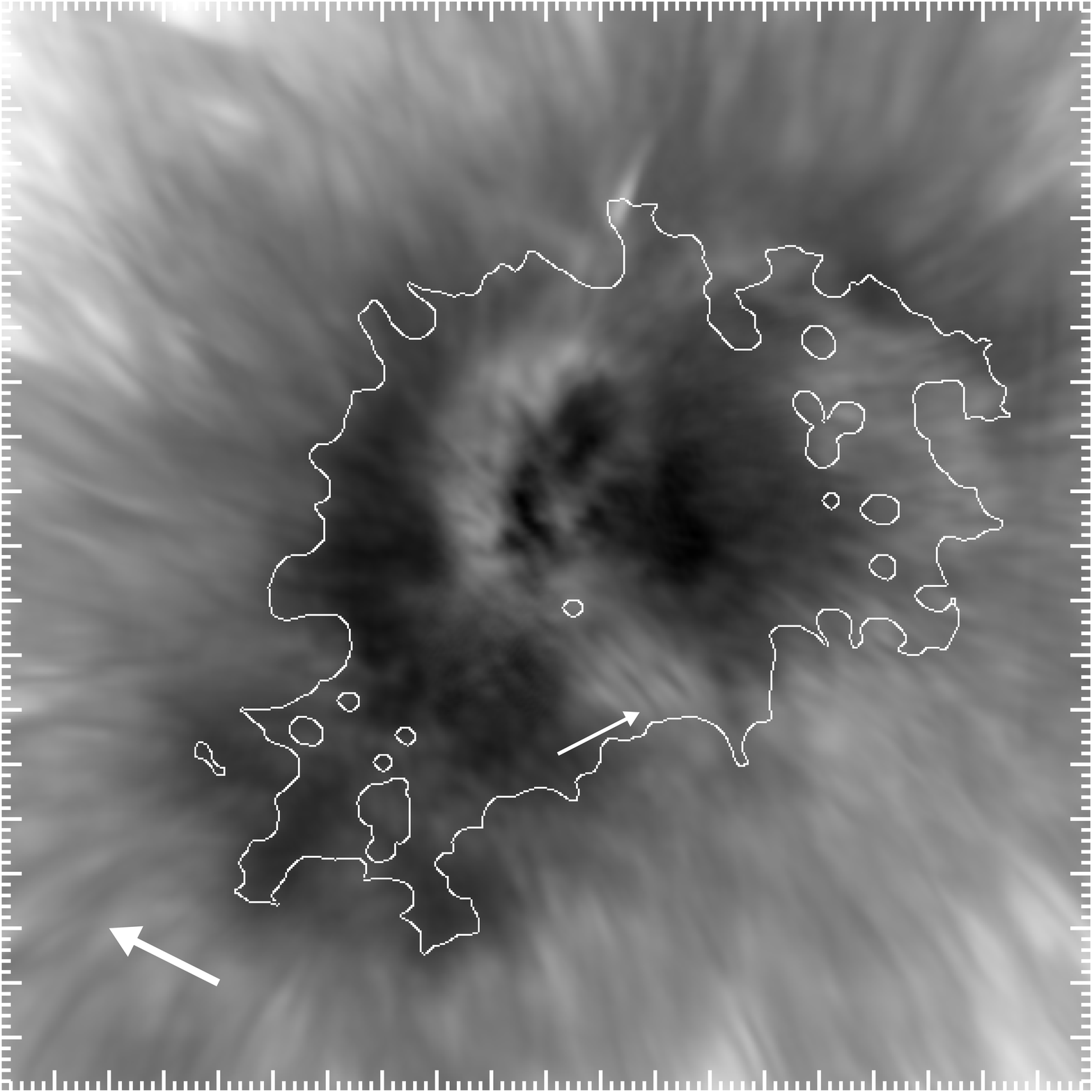}}
\resizebox{7.5cm}{!}{\includegraphics[clip=true]{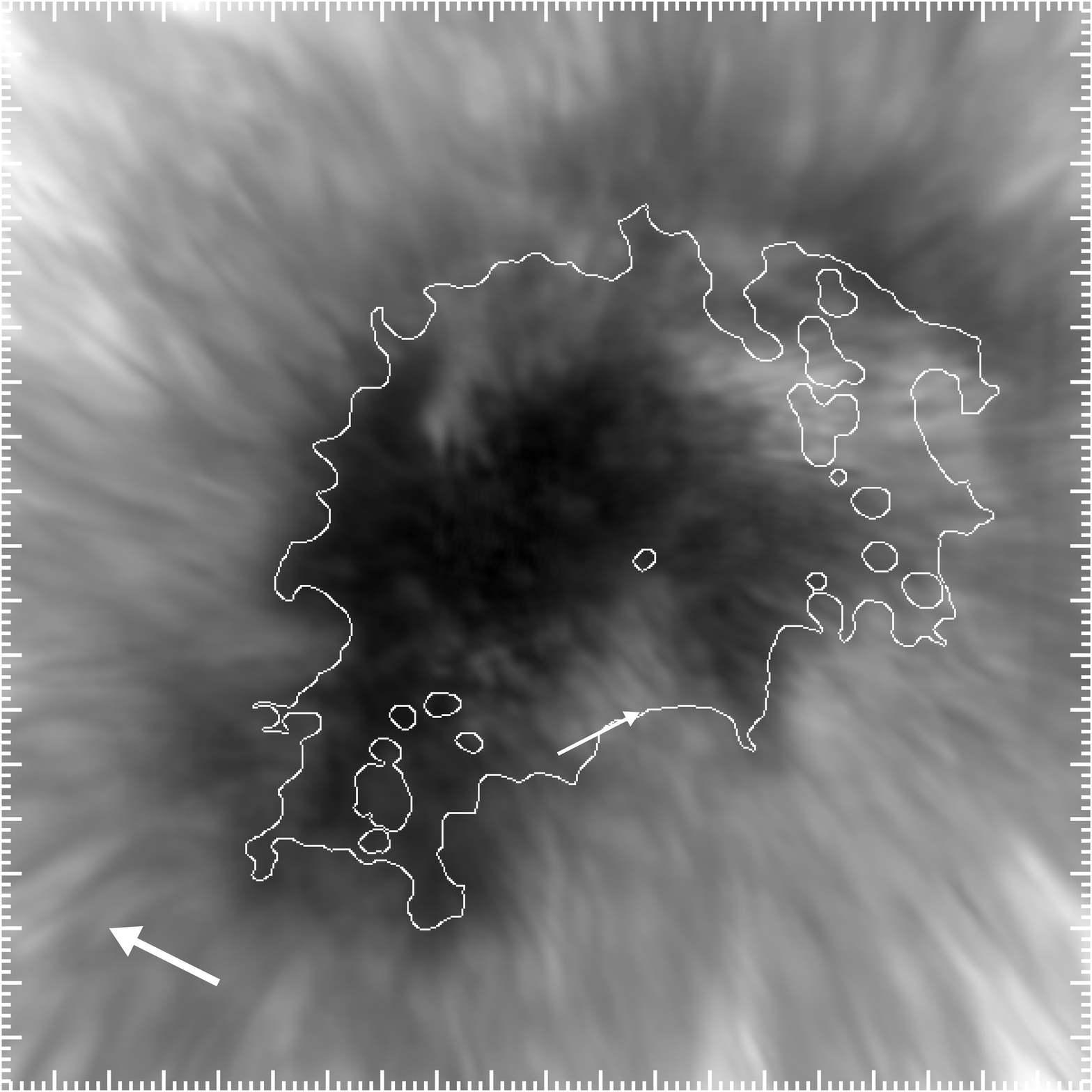}}

\resizebox{7.5cm}{!}{\includegraphics[clip=true]{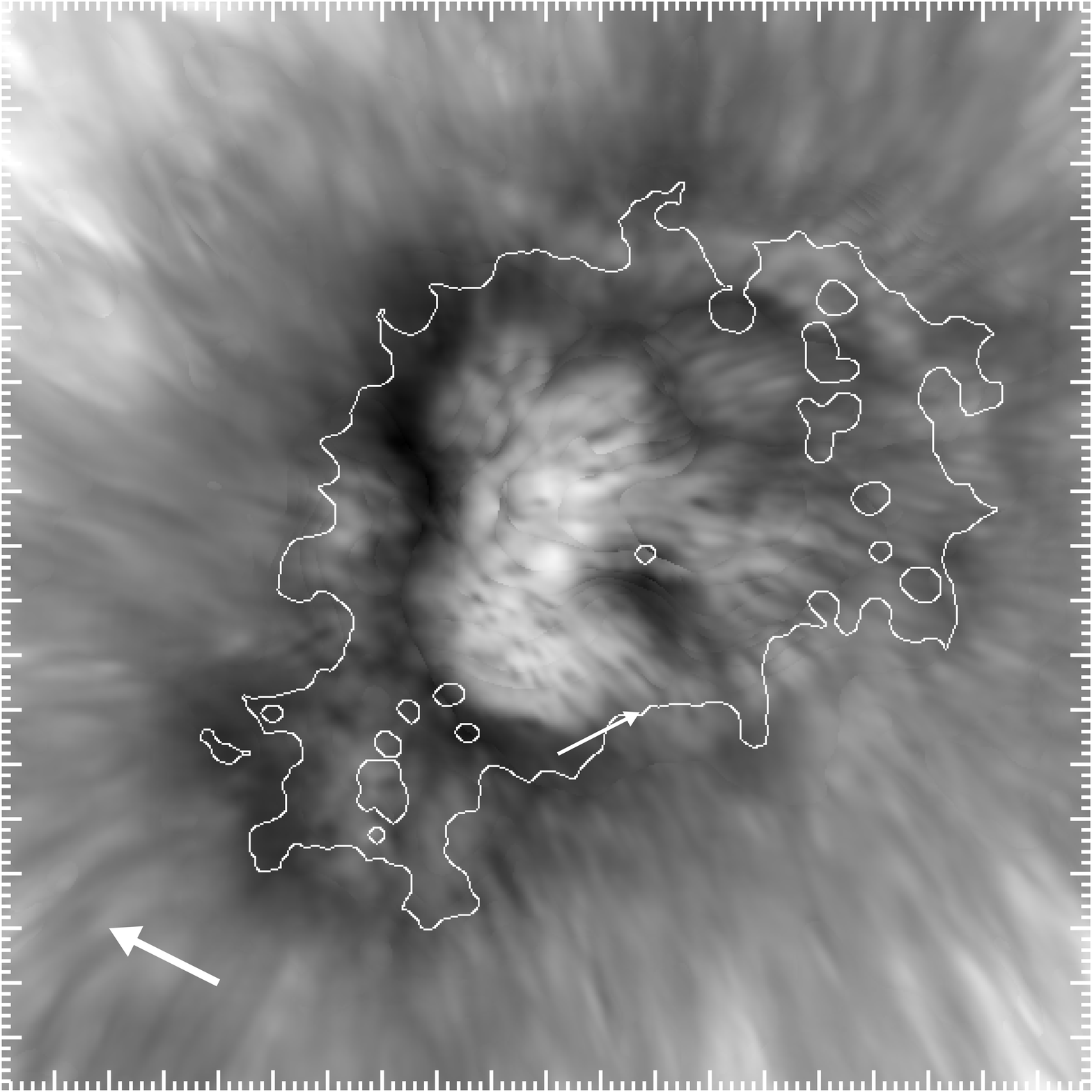}}
\resizebox{7.5cm}{!}{\includegraphics[clip=true]{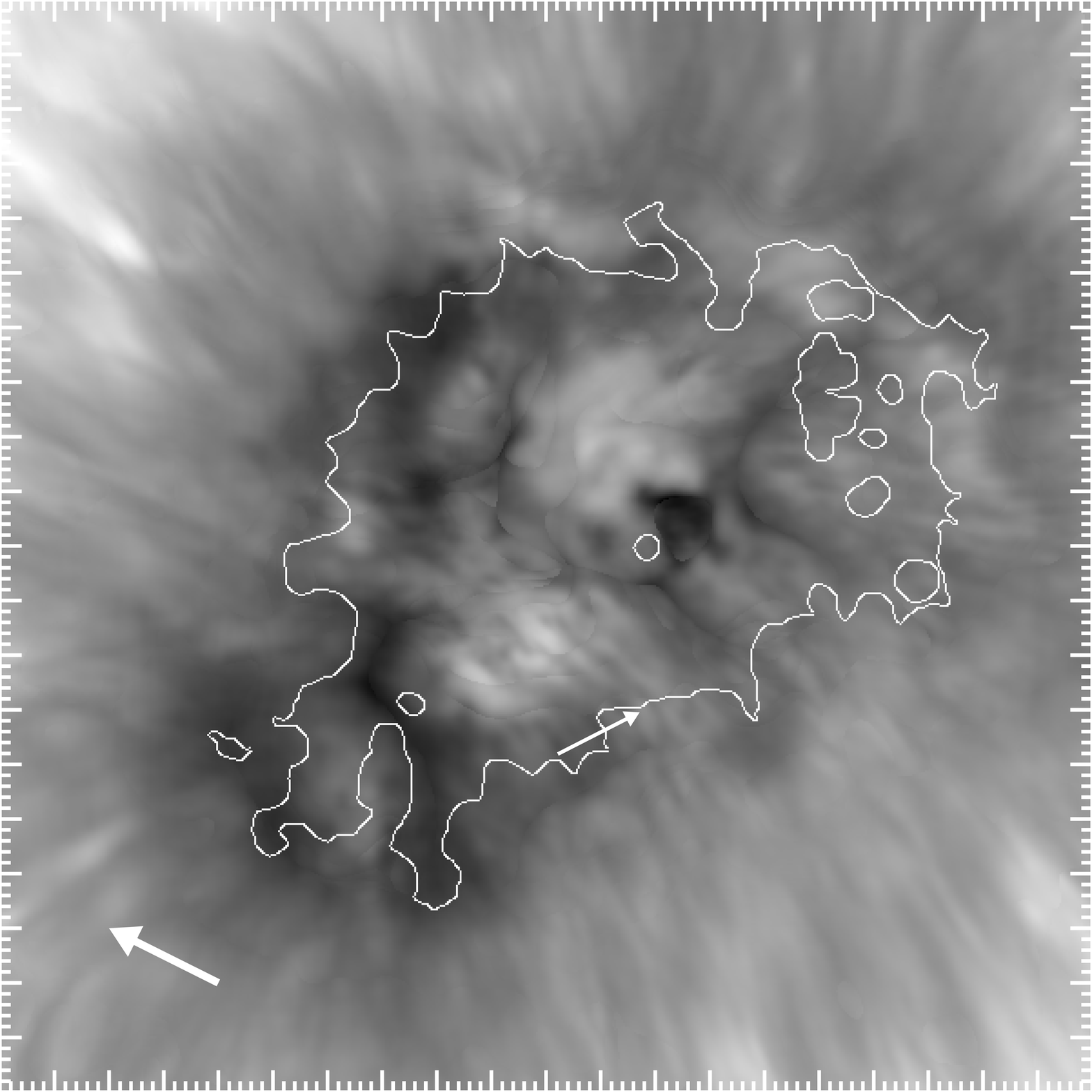}}

\resizebox{7.5cm}{!}{\includegraphics[clip=true]{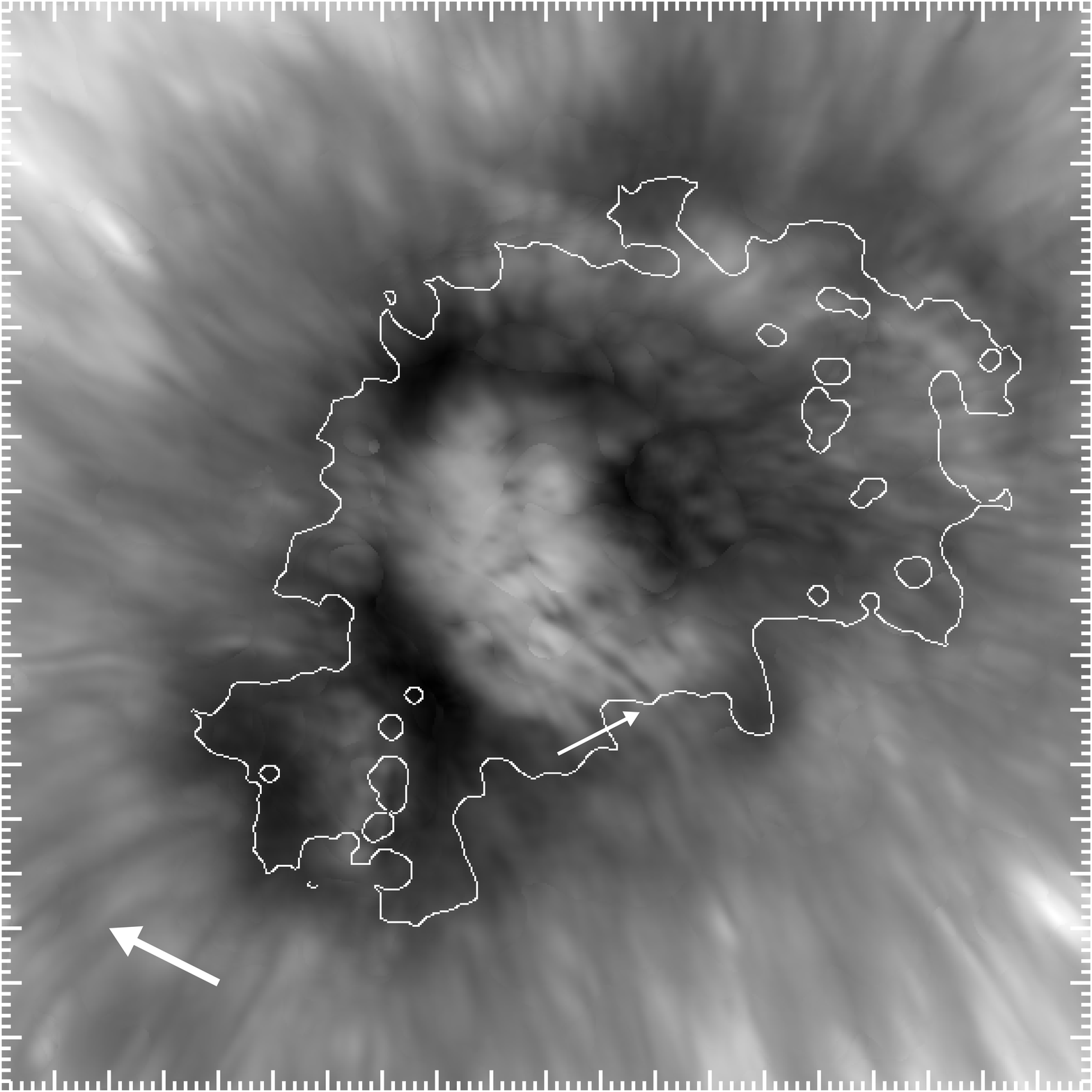}}
\resizebox{7.5cm}{!}{\includegraphics[clip=true]{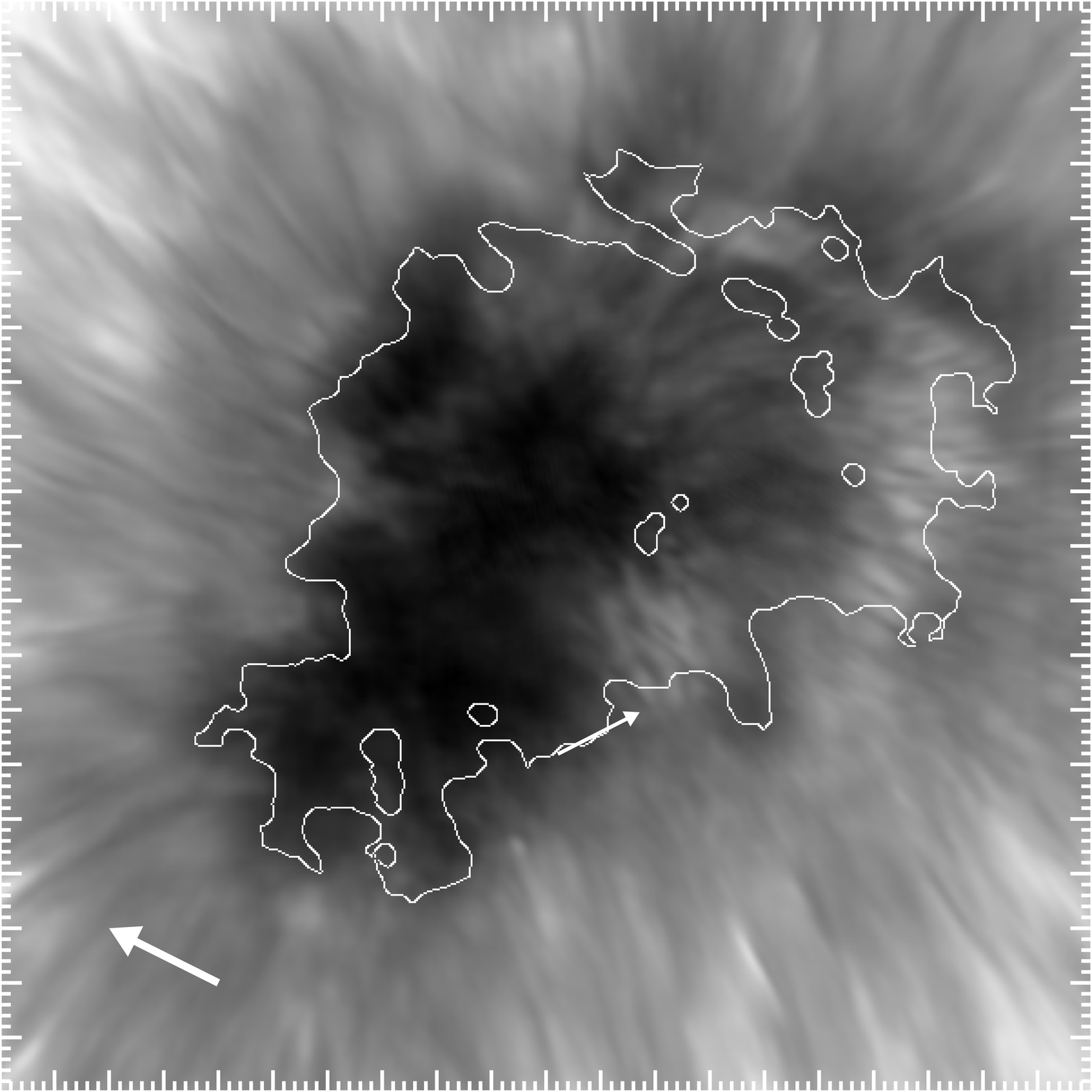}}

\caption{\footnotesize Each image in this panel is a different flash from Spot C at t=11.20, 14.42, 16.80, 19.60, 21.98, 30.52~min from the top left to the bottom right. Log-scale was used. The bottom arrow indicates the disk-centre direction. The centre arrow points to an UCF selected from the top left panel. The times t=16.80, t=19.60, and t=16.80~min, were composed using the technique described in Section~\ref{flashtracking}. The frames without composition are shown in the movie provided with the online version. Major tickmarks every 1\arcsec.}
\label{spotc}
\end{figure*}

In Fig.~\ref{subfields} different regions of interest, $3 \times 3$~arcsec wide, in the umbra of the three sunspots are shown. These sub-regions exhibit groups of UCFs that appear very similar from one flash to the next, and often to UCFs multiple flashes apart. Note that the same region can be subject to the passing of flashes within less than 3 minutes, that come from different apparent source areas. This is the case in Spot B (visible in the online time-sequence of Fig.~\ref{trails}). Although the UCFs evolve with time, their orientation appears to be always the same whenever they are visible.  Furthermore, their morphology maintains some characteristics: it is clear that the structures in the three left-most fields of the bottom row are much more similar between themselves than with those of any other row. The same can be said about most structures in any row. Still considering the bottom row as an example, one can always find a very similar UCF, in both thickness and curvature within less than one arcsec throughout 23 minutes of flashes. In the majority of cases studied here, the simplest interpretation is that there is a relatively long-lived identity to these structures, such that, in each flash, we are seeing the same canopy with some evolution. In some extreme cases this evolution can be strikingly small, as visible in the double structure of Fig.~\ref{trails}. Additionally, structures such as the one in the bottom panel of Fig.~\ref{subfields}, seem to indicate that the preserved characteristics of the atmosphere at these locations, may go beyond the magnetic field properties alone.


All the structures seem to be constituted by either a dark lane over a relatively uniform bright region or a dark fibril  bordered by two bright fibrils. The latter type is also visible over the penumbra and outside of the sunspot. In the umbra one often finds that the two bright components mix with the generally more diffuse bright background.
Fig.~\ref{subfields} also shows that most UCFs are not isolated events, but rather appear to be part of groups of UCFs where all structures have the same orientation. This is true even if one does not consider only the most visible or largest UCFs.


Fig.~\ref{trails} shows WB, Ca II H core and H-alpha images for two different flashes 2.1 minutes apart, one per column, with all the images in logarithmic scale, and the \ion{Ca}{ii}~H results composed using the technique described in Section~\ref{flashtracking}. The arrow in the figure points to a pair of UCFs that show remarkably little evolution from one flash to the next. It is difficult to believe that the source of excitation of the flash itself would be responsible for generating such similar structures so far apart. In this sense the figure is a particularly remarkable example of the relative stability (or slow evolution) shown in  the regions of interest of Fig.~\ref{subfields}. Fig.~\ref{trails} also shows a broader context to the UCFs by displaying multiple structures lit at different stages of each flash in a two minute window (over which flash tracking was applied). These appear almost 3-dimensional, and give the impression that there is a canopy fanning out from a narrow region located at the top left of the umbra of this sunspot. The triangular structure just above the pair of highlighted UCFs appears to have a footpoint (the base of the triangle) next to such narrow region. Other examples of such triangular structures were found in Sunspot A but with smaller extents. These shapes may be related to the dark regions that expand as the flash progresses, as observed in \ion{Ca}{ii}~H by \cite{2003A&A...403..277R}. 
The time-sequence associated with Fig.~\ref{trails} [see online material] was composed without flash-tracking. It samples the same field and all reconstructed frames, in log-scale, without any selection for seeing quality or interpolation in time. Despite the seeing being quite unstable (oscillating between excellent and very bad conditions) the impression of UCF permanence is striking.


In Fig.~\ref{trails} we also plot traces on top of visually selected fibrils that match the pattern of bright-dark-bright, or that are composed of a well isolated dark streak over a bright background. Traces of this type were plotted for all three sunspots and are used in Section~\ref{angle} to calculate inclination angles. When drawing these traces we did not consider fibril structures that were close, in size or intensity, to the apparent noise background. All traces were drawn from images not composed with flash tracking, and we followed a conservative approach  when it was not clear if what we were seeing was one long structure or multiple structures in close succession. We actively searched for, and found, cases where elongated and distinct structures can be aligned in such a way as to appear to be part of the same, larger, structure. However, this is not a common occurrence and, within a certain size range, it was easier to find clearly isolated patterns of fibrillar dark structures. From the traces drawn in the wideband images in the two upper panels of Fig.~\ref{trails}, we see that UCFs do not have an obvious relation to deep photospheric structures such as umbral dots. This is consistent with the results of \cite{2014ApJ...787...58Y}, where the H-alpha fine structure was found to have apparent footpoints that appeared rooted consistently in dark umbral areas as opposed to umbral dot areas.


In Fig.~\ref{spotc} we show multiple flashes where a system of UCFs with an extremely extended structure (5\farcs 0) is visible. If this is indeed one single structure, then at least some portions will be nearly horizontal and it is very unlikely that the inclination angle will be constant throughout its full extent. UCFs occasionally appear to have forked foot-points and this structure is a good example. In the bottom left panel it is clear that the structure extends from close to the centre of the umbra to over the penumbra. It is roughly aligned with the continuum penumbral filaments (like all UCFs when extended radially) and it slightly changes direction when crossing the umbra-penumbra boundary to follow the other fibrillar structures above the sunspot's penumbra. This slight change of direction is visible for other, if smaller, structures. When observing the fibrillar structures above a sunspot penumbra, such as the line-core panels from Fig.~\ref{fig:intro}, there is an impression that the ensemble of the structures is inclined, with the lower height being in the umbral boundary. Since the UCF bends to follow the same orientation as the fibrils over the penumbra, then it might simply be bending more towards the vertical. This ``bending'' is very visible in the time-series associated with Fig.~\ref{spotc} [see online material], whenever an umbral flash propagates through the umbra-penumbra boundary. This time sequence is also composed without flash-tracking but log-scale was used.

In the two bottom panels of the left row of Fig.~\ref{spotc}, many small, yet filamentary structures are clearly visible above the background of two different powerful flashes. These are illuminated in the early stages of the flashes (first 34 seconds) and appear very close to the centre of the umbra. Due to their proximity with the darkest part of the umbra, when compared with the longer structure, it is tempting to view these as the foot-points of structures with smaller angles to the vertical. These UCFs have the same orientation as the other longer structures, including the extremely extended structure visible in the same frames. 

\subsection{Angle versus formation range}
\label{angle}

The traces drawn as described in the previous section allow us to measure a horizontal projection in km. This is performed for several flashes and for those structures that meet the criteria used to draw the traces. 

To compute the vertical extent (and therefore the angle with the vertical) we either need many statistics on UCFs at very different $\mu$ (which we do not have), or a height range over which these structures are formed. If one assumes that the dark structures are formed at the same height as the flashes, then the latter is provided by the \ion{Ca}{ii}~H grain formation heights of \cite{2010ApJ...722..888B}, as all the observed intensity enhancement during the flash comes from those spectral features.  We note that, while we are observing with a 1.1~\AA\ interference filter, the umbral flashes are chromospheric phenomena and no structures are observed in the inner wing (upper photosphere) of  \ion{Ca}{ii}~H  in \cite{2013A&A...557A...5H}. 

\cite{2010ApJ...722..888B}  obtain peaks for the umbral flash formation between 1,100~km and 1,300~km above the photosphere. In addition, from the plots of their contribution functions, the ranges that the flash illuminates at any given point appear to have a typical extent of approximately 300~km and cover a range of about 500~km while they progress upwards. In Fig.~7 of \cite{2010ApJ...722..888B}, where the late stage of a flash is depicted, the contribution function appears to have a faint but noticeable tail that extends from 1,500~km to 2,000~km, bringing the total height range that contributes to emission to 1,000~km.

\begin{figure*}[!htb]
  \centering
\resizebox{12cm}{!}{\includegraphics[clip=true]{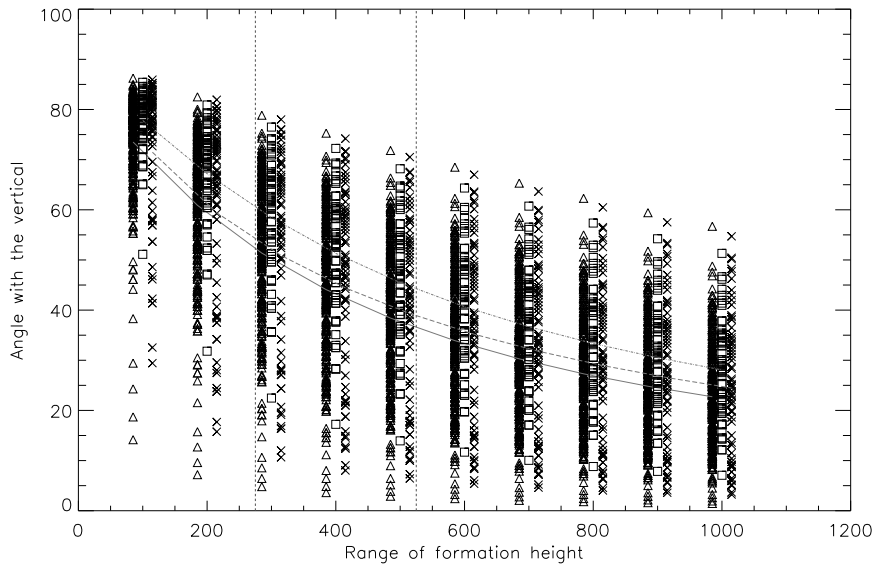}}

\caption{\footnotesize Angle with the vertical versus formation-height range for all structures measured and all spots, with triangles standing for structures measured in Spot A, squares for Spot B and crosses for Spot C. The symbols are shifted 15~km left for Spot A and 15~km right for Spot C for visibility. The inclination angle is plotted for each individual structure as a function of potential formation range. The vertical lines delimit the range of formation heights we expect from \cite{2010ApJ...722..888B}. The solid line (lowest) connects the average angles for Spot A, the dashed line connects the average angles for Spot C (middle line), and the dash-dotted line the average for Spot B.} 
\label{angles}
\end{figure*}

We plot, in Fig.~\ref{angles}, the inclination angles (from the vertical) by taking the arc-tangent of the ratio between the horizontal extent and the formation range, for an interval of formation ranges. The horizontal extents were conservatively corrected for $\mu$ angle by assuming that we always see UCFs from the direction that would most increase their apparent horizontal extent. The interval plotted is 100 to 1000~km to include all possible ranges mentioned above. It provides a visual impression of the potential inclination of the structures in a wider range of vertical formation ranges. From this figure we see that the theoretical formation-height ranges of \cite{2010ApJ...722..888B} would imply large angles from the vertical for most structures traced in our paper, with averages of 50 and 60 degrees from the vertical for the 500~km  and 300~km range cases, respectively. These plots do not include extremely large structures such as the one shown in Fig.~\ref{spotc}, as we conservatively use only frames not composed with the technique described in Section~\ref{flashtracking}.  This latter structure has a maximum extension of 3800~km, which would imply an inclination of 75 degrees from the vertical assuming a vertical extent of 1000~km. As mentioned above, the structure appears to change direction close to the umbra/penumbra boundary which might also denote a change in inclination.

\subsection{Similarities with H-alpha}
\label{similarities}
Any comparison between \ion{Ca}{ii}~H and H-alpha in the umbra is affected by the fact that we are observing two dynamic phenomena.  H-alpha dark structures alternatively increase and decrease their horizontal extent as observed by \cite{2013ApJ...776...56R}. If the same behaviour is present in the structures traced by UCFs, we will be unable to observe it since they are revealed only briefly as the umbral flash moves radially outwards. This limits a detailed comparison, not only in terms of dynamics but also in horizontal extent and correspondence.

Despite this difficulty, correspondence between many of the identified UCFs and H-alpha dark structures is clearly visible, as shown in Fig.~\ref{trails}, with examples of such correspondence labelled with the numbers 4,5,6,9,14,15,16,17,18, 20, and 21. This supports the argument made in \cite{2013ApJ...776...56R} that the UCFs observed in \ion{Ca}{ii}~H by \cite{2009ApJ...696.1683S} and \cite{2013A&A...557A...5H} are counterparts of the H-alpha short dynamic fibrils. However, there are several UCFs with no visible correspondence in H-alpha (structures numbered 1,2,3,7,8,10,11, and 19). The horizontal extents for structures with a correspondence are almost always different.  In \cite{2013ApJ...776...56R}, dynamic structures are observed in \ion{Ca}{ii}~8542 that appear to have a strong relation with the H-alpha short dynamic fibrils. These are located in a region of great complexity that appears to be a mixture of small abnormal granulation, penumbra and umbra. The authors find that the extents of the H-alpha structures are larger than the structures observed in \ion{Ca}{ii}~8542. Here we find \ion{Ca}{ii}~H structures that are larger than their apparent H-alpha counterpart (e.g.~15,18,21). This would make sense if the sequential increase in opacity (8542, H-alpha, \ion{Ca}{ii}~H) translates to sampling these structures sequentially higher while still capturing the foot-points for all three lines.

We find some apparent differences between the morphology of the UCFs and the candidate H-alpha counterparts. UCFs have a marked bright-dark-bright structure and appear generally slender. For the latter, both the difference in resolution between the two wavelengths and the scattering nature of H-alpha are likely to play a role. In the centre of Fig.~\ref{trails}  we see a triangular UCF, marked with the number 16, that narrows from the upper left corner to the lower right corner. This is suggestive of the ``cone-like'' appearance of the spikes from \cite{2014ApJ...787...58Y}. However, the H-alpha counterpart in these observations, does not show the same shape. 

The widths and transversal profiles of the structures tend to be similar to that highlighted in Fig.~7 of  \cite{2013A&A...557A...5H}. The width of the central dark filament is approximately 0\farcs15 and indicates that it is not completely resolved in most structures.
 
\subsection{Origin of the dark streaks} 

We find that the bright-dark-bright pattern in the fibrils observed in the umbra, is very similar to that observed in the fibrils above the penumbra and even the fibrils observed outside of the sunspot. For example, one can see in the very clear structures of Fig.~\ref{trails} what seems to be two bright components next to most of the dark streaks. We therefore propose that the cores of UCFs may be dark for the same reason that the fibrils visible in \ion{Ca}{ii} lines have dark streaks in general. Recently, \cite{2014ApJ...788..183B} proposed that the dark lanes observed in the \ion{Ca}{ii}~8542 fibrils, are due to increased opacity due to strong flows in the centre of the structures. This is similar to the model of \cite{2008A&A...488..749R}, originally put forward to explain photospheric penumbral dark-cores \citep{2002Natur.420..151S}. We note that, if strong flows are responsible for the dark streaks in UCFS, this would fit perfectly with the scenario of two components shifted in velocity, with one component down-flowing and the other either shocked or up-flowing, as first put forward in \cite{2000Sci...288.1398S}. Both in \cite{2000Sci...288.1398S}  and \cite{2014ApJ...788..183B} the inverse Evershed effect is proposed as a source for the flows.

 In \cite{2013ApJ...776...56R}, up-flows and down-flows in the blue and red wings (respectively) of the \ion{Ca}{ii}~8542 line are clearly visible and have a direct relation with the movements of the H-alpha short dynamic fibrils. Asymmetric shock-front propagation, present in an atmosphere with only minor inhomogeneities, as seen in the 2D simulations of \cite{2011ApJ...743..142H}, is proposed by \cite{2013ApJ...776...56R} as the explanation for short dynamic fibrils. This may also play a role in UCF formation. In \cite{2013ApJ...776...56R} there seems to be a direct relation between length, lifetime and the magnetic field inclination (as measured in 8542) of the short dynamic fibrils, which is consistent with the simulated properties of \cite{2011ApJ...743..142H}. 

While shocks are clearly at play, it is hard to imagine the above mechanism as the sole responsible for the larger and more complex UCFs observed in his paper, such as the one shown in Fig.~\ref{spotc}. Due to this and the similarity of umbral structures with those outside the umbra, we propose that, for the structures seen in \ion{Ca}{ii}~H, steady flow inhomogeneities may play a role in creating the intensity inhomogeneities when the shocks pass through.

\section{Concluding remarks}

In this work we show that some of the individual UCFs, not just groups of similar UCFs in approximately the same location, are stable for at least the period between two flashes, and are well-delineated structures present in multiple independent frames. This excludes them from being ephemeral manifestations of shocked material or mere noise artifacts in an otherwise homogeneous and vertical chromosphere. Keeping in mind that there are no two umbral flashes alike, (as previously well shown in \cite{2003A&A...403..277R}) the probability that two very different flashes, minutes apart, would randomly produce visually horizontal structures with the degree of similarity shown in this paper, and at the same locations, in a layer of the atmosphere that would otherwise be nearly  homogeneous and vertical, is indeed very unlikely. Further supporting this statement is the stability of the groups of fibrils themselves, always showing some evolution but visible over more than two flashes for three different sunspots. The latter adds to the system reported to appear in multiple flashes by \cite{2009ApJ...696.1683S}.

\cite{2013A&A...556A.115D} found that umbral flashes appear to arise mainly from changes in temperature and velocity and not so much from changes in the magnetic field. This inferred stability of the magnetic field between flashes is compatible with our observed relative stability of the UCFs between flashes. 

We observe many UCFs as composed by two bright components (brighter than the surroundings) and a central dark structure, rather than a dark structure over a bright background. Due to such pattern being similar to that of the \ion{Ca}{ii}~H fibrils visible outside of the sunspot, we have put forward the idea that the same mechanism is behind the intensity inhomogeneity visible in both cases. We also propose that such a mechanism may be the general flow structure, even if shocks are clearly responsible for the visibility of the UCFs. We are unable to completely exclude that these central dark streaks are not absorption features formed above the flash formation heights.

This work, together with that of \cite{2009ApJ...696.1683S}, shows that UCFs are common in sunspots and that one should expect to find them in all observations of sufficiently high spatial resolution. In all four spots where UCFs have been observed, the implied inclination angles for some structures, when using formation ranges from simulations such as \cite{2010ApJ...722..888B}, are close to horizontal, even for the largest vertical formation ranges. Better theoretical constraints on the formation height of the \ion{Ca}{ii}~H bright component during flashes are required. There is a lack of information regarding the chromospheric umbral atmospheres. Theoretical works like the one of \cite{2010ApJ...722..888B} assume atmospheres with a vertical magnetic field configuration. This reduces the reliability of the angles that we estimate since the flashes could potentially propagate further in a wave-guide that is more inclined from the vertical. Future simulations could focus on the formation heights of the \ion{Ca}{ii}~H flashes in inclined magnetic field atmospheres. Alternatively, an observational study of these structures with very high statistics for different viewing angles, could provide better constraints on inclinations. Determining the exact relation between UCFs and the short dynamic fibrils of \cite{2013ApJ...776...56R} (or the similar spikes from \cite{2014ApJ...787...58Y}) would help towards this purpose. To this end, we find that, while there is a correspondence between UCFs and dark structures in H-alpha, there is no simple one to one relation. 

Other works that hint at a chromosphere in the umbra that could harbour strongly inclining fields include \cite{2014ApJ...788..183B} and \citep{2013A&A...553A..63S}. The former find most of the observable \ion{Ca}{ii}~8542 super-penumbral canopy to connect back with the photosphere close to the sunspot, and the latter find the presence of opposite polarity patches in the penumbra itself, including the umbra/penumbra boundary.  

We find both extremely elongated UCFs extending into the penumbra, and very small UCFs in groups that are consistent with what could be foot-points of nearly vertical structures. The presence of these two extremes indicates a variable inclination magnetic field and a very in-homogeneous umbra. Furthermore, the extremely large UCFs change direction when going over to the penumbra, consistent with mostly horizontal structures changing their inclination with the vertical. 

Based on these observations, the likelihood that the fibrillar structures do trace the magnetic field, and the fact that field lines tend to fan out very quickly in strong magnetic fields, we speculate that the magnetic field structure in a sunspot's chromosphere does indeed ``fan out" rather rapidly. This creates magnetic-dominated structures at all inclinations, and the more elongated structures observed here are the lower-height observable manifestations of very inclined magnetic field lines pushed down by the remaining, more vertical, magnetic field. Our scenario would be similar to classic ``wine-glass'' models, just at a larger scale and with the ``base'' of the glass forming surprisingly early in the umbra. This would still be fully compatible with the observations and simulations of the saw-toothed pattern discussed in the introduction, as well as with waves reaching the transition region (recently detected by \cite{2014ApJ...786..137T}). Still trying to make sense of the topology of the magnetic field, we cannot exclude a scenario similar to that proposed by \cite{2011ApJ...742..119R} to explain the canopies  observed in a network region, where two components of the chromospheric field would occur interlaced, one vertical and another one more horizontal, with the horizontal component carrying a small fraction of the total magnetic flux.

The main difficulty in studying the dynamics of the UCFs is that these are only observable when they are illuminated by the umbral flash. This, in turn, should provide different height samplings when sunspot observations in this line become routine at 0\farcs10 or better (as well as information about dynamics). If the structures are visible but just very faint during the dark phase of the flashes, then the greater light-gathering capacity provided by larger aperture telescope such as the DKIST and the planned EST will allow us to observe these structures even when the flash is not present, thus giving us access to detailed UCF dynamics. Alternatively, longer exposure times could be attempted. However, in the blue, smearing can increase very fast with even small increments in exposure so the observer needs to be aware of that when attempting this exercise in excellent seeing conditions. A negative detection in excellent seeing with high exposure times during the dark phase of the flash, followed by positive detection in the bright phase, would reinforce the view that shocks are responsible for both the emission and the intensity inhomogeneity due to shock front asymmetry.

\begin{acknowledgements}
We would like to thank Peter S\"{u}tterlin, Johanna Vos, and Peter Halpin for assisting with the observations. We would also like to acknowledge support from Robert Ryans, Chris Smith, David Malone, and Gabriele Pierantoni with computing infrastructure. We thank Hector Socas-Navarro for valuable discussions. We further thank the anonymous referee for comments and suggestions that improved the manuscript. This work made use of an IDL port of Robert Shine's routines made by Tom Berger. This work is supported by the Science and Technology Facilities Council. Eamon Scullion is a Government of Ireland Post-doctoral Research Fellow support by the Irish Research Council. This research was supported by the SOLARNET project (www.solarnet-east.eu), funded by the European Commission´s FP7 Capacities Program under the Grant Agreement 312495. The Swedish 1-m Solar Telescope is operated on La Palma by the Institute for Solar Physics of the Royal Swedish Academy of Sciences in the Spanish Observatorio del Roque de los Muchachos of the Instituto de Astrof\'isica de Canarias. 
\end{acknowledgements}

\bibliographystyle{aa}

\end{document}